\documentclass[final,5p,times,twocolumn]{elsarticle}

\usepackage{amsmath}
\usepackage{array, longtable, tabularx}
\usepackage{comment}
\usepackage{hyperref}
\usepackage{placeins}
\usepackage{amssymb}
\hypersetup{ 
 colorlinks=true, 
 citecolor=black, 
 filecolor=black, 
 linkcolor=black, 
 pdfauthor=author,
 } 
\usepackage{enumitem} 

\journal{}
 
\begin{document}
\begin{frontmatter}

\title{
A Segmented Total Energy Detector (sTED) optimized for (n,$\gamma$) cross-section measurements at n\_TOF EAR2\\
}
 
\author[1]{\small{V.~Alcayne\corref{cor1}}}\ead{victor.alcayne@ciemat.es}
\author[1]{D.~Cano-Ott}
\author[1]{J. Garcia}
\author[1]{E.~Gonz\'{a}lez-Romero}
\author[1]{T.~Mart\'{i}nez}
\author[1]{A.~P\'{e}rez de Rada}
\author[1]{J.~Plaza}
\author[1]{A.~S\'{a}nchez-Caballero}
\author[2]{J.~Balibrea-Correa}
\author[2]{C.~Domingo-Pardo}
% ...
%

\address[1]{\scriptsize{Centro de Investigaciones Energ\'{e}ticas Medioambientales y Tecnol\'{o}gicas (CIEMAT), Spain}} %
\address[2]{Instituto de F\'{\i}sica Corpuscular, CSIC - Universidad de Valencia, Spain} %
% ...
%

\cortext[100]{Corresponding author at CIEMAT, Complutense 40, 28040 Madrid, Spain. Tel: +34 913466614.}

\date{\today}


\begin{keyword}
Neutron capture \sep PHWT \sep Scintillation detectors \sep Monte Carlo simulation
\end{keyword}

\end{frontmatter}
\section*{Abstract}

The neutron time-of-ﬂight facility n\_TOF at CERN is a spallation source dedicated to measurements of neutron-induced reaction cross-sections of interest in nuclear technologies, astrophysics, and other applications. Since 2014, Experimental ARea 2 (EAR2) is operational and delivers a neutron fluence of $\sim$4$\cdot$10$\mathrm{^{7}}$ neutrons per nominal proton pulse, which is $\sim$50 times higher than the one of Experimental ARea 1 (EAR1) of $\sim$8$\cdot$10$\mathrm{^{5}}$ neutrons per pulse. The high neutron flux at EAR2 results in high counting rates in the detectors that challenged the previously existing capture detection systems. For this reason, a Segmented Total Energy Detector (sTED) has been developed to overcome the limitations in the detector's response, by reducing the active volume per module and by using a photo-multiplier (PMT) optimized for high counting rates. This paper presents the main characteristics of the sTED, including energy and time resolution, response to $\gamma$-rays, and provides as well details of the use of the Pulse Height Weighting Technique (PHWT) with this detector. The sTED has been validated to perform neutron-capture cross-section measurements in EAR2 in the neutron energy range from thermal up to at least 400 keV. The detector has already been successfully used in several measurements at n\_TOF EAR2.


\section{Introduction}
\begin{figure}[!ht]
\centering
\includegraphics[width=0.49\textwidth]{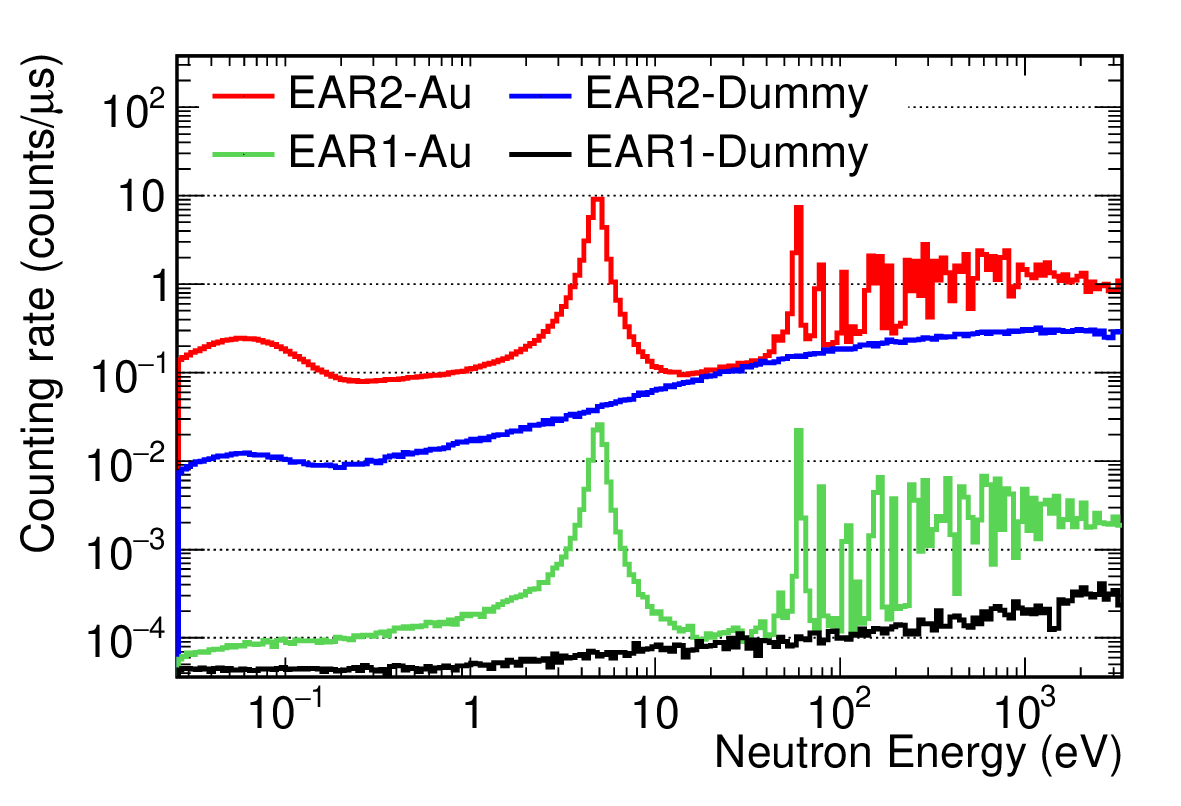}
\caption{Counting rates obtained as a function of the neutron energy in the experimental EAR1 (EAR1-Au) and EAR2 (EAR2-Au) for a BICRON detector with a threshold of 0.15 MeV. The detectors are located at 10 cm from a $^{197}$Au sample of 2 cm in diameter and 100 $\mu$m thickness. The counting rates of the background obtained when measuring the dummy, hereafter a setup equal to the one of $^{197}$Au but without the $^{197}$Au sample, are also presented for the EAR1 (EAR1-Dummy) and the EAR2 (EAR2-Dummy).}
\label{fig:CR_EAR1_EAR2} 
\end{figure}

The neutron time-of-flight facility n\_TOF at CERN is focused on performing measurements of neutron-induced reaction cross-sections of interest to nuclear technologies, astrophysics, and other applications. The facility uses as a neutron source a massive lead spallation target coupled to the CERN-PS 20 GeV/c proton beam \cite{Esposito_target_23} and is endowed with three experimental areas: Experimental ARea 1 (EAR1) with $\sim$8$\cdot$10$\mathrm{^{5}}$ neutrons per nominal pulse of $\sim$7$\cdot$10$\mathrm{^{12}}$ protons, located at $\sim$185 m horizontally from the spallation target \cite{Guerrero_13}, Experimental ARea 2 (EAR2) with $\sim$4$\cdot$10$\mathrm{^{7}}$ neutrons per pulse, located vertically at $\sim$20 m from the spallation target \cite{Weiss_15}, and the recent NEAR station with $\sim$4$\cdot$10$\mathrm{^{9}}$ neutrons per pulse, at $\sim$3 m from the target currently under commissioning \cite{Ferrarri_23,Gervino_22}. EAR2 was constructed to carry out challenging cross-section measurements with low mass samples, reactions with small cross-sections and/or highly radioactive samples \cite{Barbagallo_16,Sabate_17,Damone_18,Stamatopoulos_20,Alcayne_19}. As evident from the numbers above the neutron flux in EAR2 is $\sim$50 times higher than in EAR1 and the neutrons take $\sim$10 times less time to arrive at the experimental area. As a consequence, the signal-to-background ratio is increased by a factor of $\sim$500 when considering the constant room background or the radioactivity of the samples. Accordingly, the counting rate in the detectors is also increased by approximately the same factor, as presented in Fig. \ref{fig:CR_EAR1_EAR2}, which implies considerable experimental challenges.

Capture cross-section measurements with C$_6$D$_6$ detectors have been performed successfully at n\_TOF EAR1 for about 20 years \cite{Guerrero_14,Gunsinsg_17}. In most cases, the analysis of the C$_6$D$_6$ detector data was done by applying the Pulse Height Weighting Technique (PHWT) \cite{Macklin_67,Tain_04,Borella_07}, which allows the C$_6$D$_6$ to mimic the behavior of an ideal Total Energy Detector (TED) \cite{Moxon_63}. The measurements were mainly performed with commercial BICRON detectors (0.621 liters of C$_6$D$_6$) \cite{Plag_03} and self-made and customized detectors with carbon-fiber housing (1.0 liters of C$_6$D$_6$) \cite{Mastinu_13}. The photomultipliers of the two detectors were not optimized for the high count rates of EAR2.

Two capture cross-section measurements have been carried out in EAR2 \cite{Alcayne_19,Oprea_20} with these detectors and considerable pile-up effects and gain shifts were observed in the data due to the challenging conditions of this area \cite{Alcayne_22}. These effects increased with the neutron energy (i.e. at shorter times of flight) and required the introduction of considerable corrections in the capture cross-section data analysis. These corrections made almost impossible to perform capture cross section measurements above a few keV at EAR2 with these detectors. To overcome these limitations, a Segmented Total Energy Detector (sTED) \cite{Alcayne_23} has been developed. It consists of an array of small active volume C$_6$D$_6$ modules coupled to photomultipliers optimized for high counting rates applications.

This paper discusses the performance of the large volume C$_6$D$_6$ detectors at EAR2 in Section \ref{section:Limitations}, the properties of the new sTED detector in Section \ref{section:Detector} and the performance at n\_TOF EAR2 of this new detector in Section \ref{section:Validation}. The summary and conclusions of this work are presented in Section \ref{section:Conclusions}.

 \section{Performance of C$_6$D$_6$ detectors at EAR2}\label{section:Limitations}
As mentioned previously, the BICRON detectors and the self-made detectors with carbon-fiber housing, have been commonly used at EAR1, but stand very high counting rates in capture measurements at EAR2. Hence, the measured data suffer from the effects described in the two following subsections.

 \subsection{Pile-up effects}\label{section:PileUp}
\begin{figure}[!ht]
\centering
\includegraphics[width=0.49\textwidth]{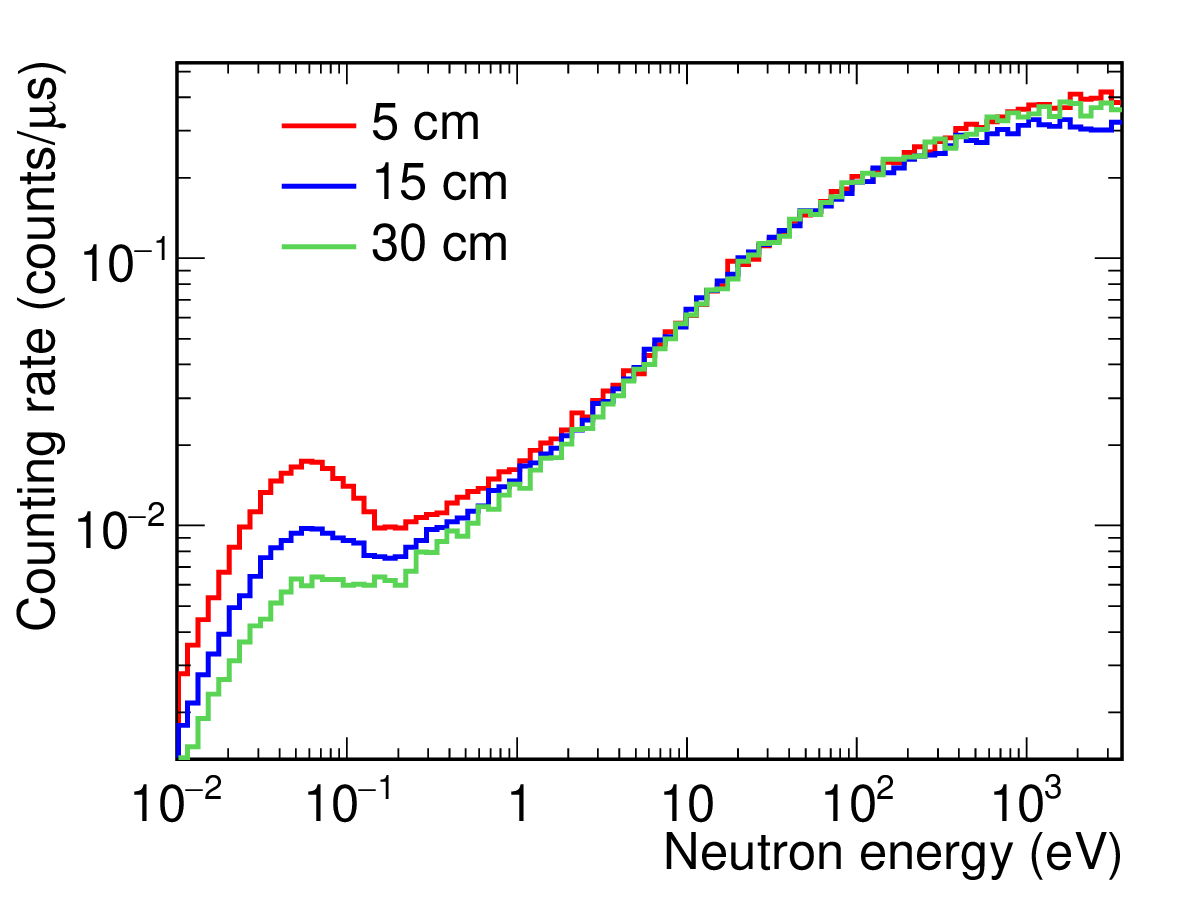}
\caption{Counting rates obtained for a BICRON detector as a function of the neutron energy with a 0.15 MeV deposited energy threshold in a beam-on measurement without any sample in place at the nominal proton intensity. The measurements are performed with the detector at the same distance from the spallation target but at three different distances (5, 15 and 30 cm) from the center of the beam.}
\label{fig:VariousDistances} 
\end{figure}

 As shown in Fig. \ref{fig:CR_EAR1_EAR2} for a BICRON detector at 10 cm from a 100 $\mu$m thick and 2 cm diameter gold sample the counting rate reaches up to 10 counts per $\mu$s. Such a high counting rate with the $\sim$10 ns Full Width at Half Maximum (FWHM) signals of the used large volume C$_6$D$_6$ detectors leads to significant pile-up effects. Pulse shape fitting can be used for reconstructing piled up signals \cite{Guerrero_08}, but even with this technique $\sim$25\% of the signals are lost at a counting rate of 10 counts per $\mu$s \cite{Alcayne_22}. One possible solution for reducing the pile-up is to move the C$_6$D$_6$ detectors away from the sample, thus lowering the efficiency and therefore the count rate. However, measurements performed in EAR2 with BICRON detectors have shown that the neutron beam-related background (i.e. counts in the detector without any sample) is almost constant at different distances from the center of the beam for C$_6$D$_6$ detectors in a wide range of neutron energies, as presented in Fig. \ref{fig:VariousDistances}. Therefore, moving the detectors further away from the sample decreases the signal to beam-related background. Since this background of EAR2 is one of the main limitations for performing capture measurements there \cite{Alcayne_22,Mendoza_23}, it does not help much to move the C$_6$D$_6$ detectors further away from the sample.
 \subsection{Gain shift effects}\label{section:GAinShift}

 At n\_TOF EAR2 three different types of gain-shift effects have been observed, which are described in the following list:
\begin{itemize}
 \item Gain shifts due to high constant counting rates appear when the counting rate in the detector increases from one constant value to a higher constant value. This gain shift has been observed and characterized in measurements performed with high activity $\gamma$-ray calibration sources. Fig. \ref{fig:GainCRL6D6} depicts an example of this effect for a carbon-fiber housing detector: the pulse height spectra in the detector recorded with strong $^{137}$Cs and $^{88}$Y sources placed together at different distances (i.e different counting rates) is shifted. The gain decreases with higher counting rates. As a consequence, the energy calibration would be modified. A similar but smaller effect was also observed in the BICRON detectors. 
\begin{figure}[!ht] 
\centering
\includegraphics[width=0.49\textwidth]{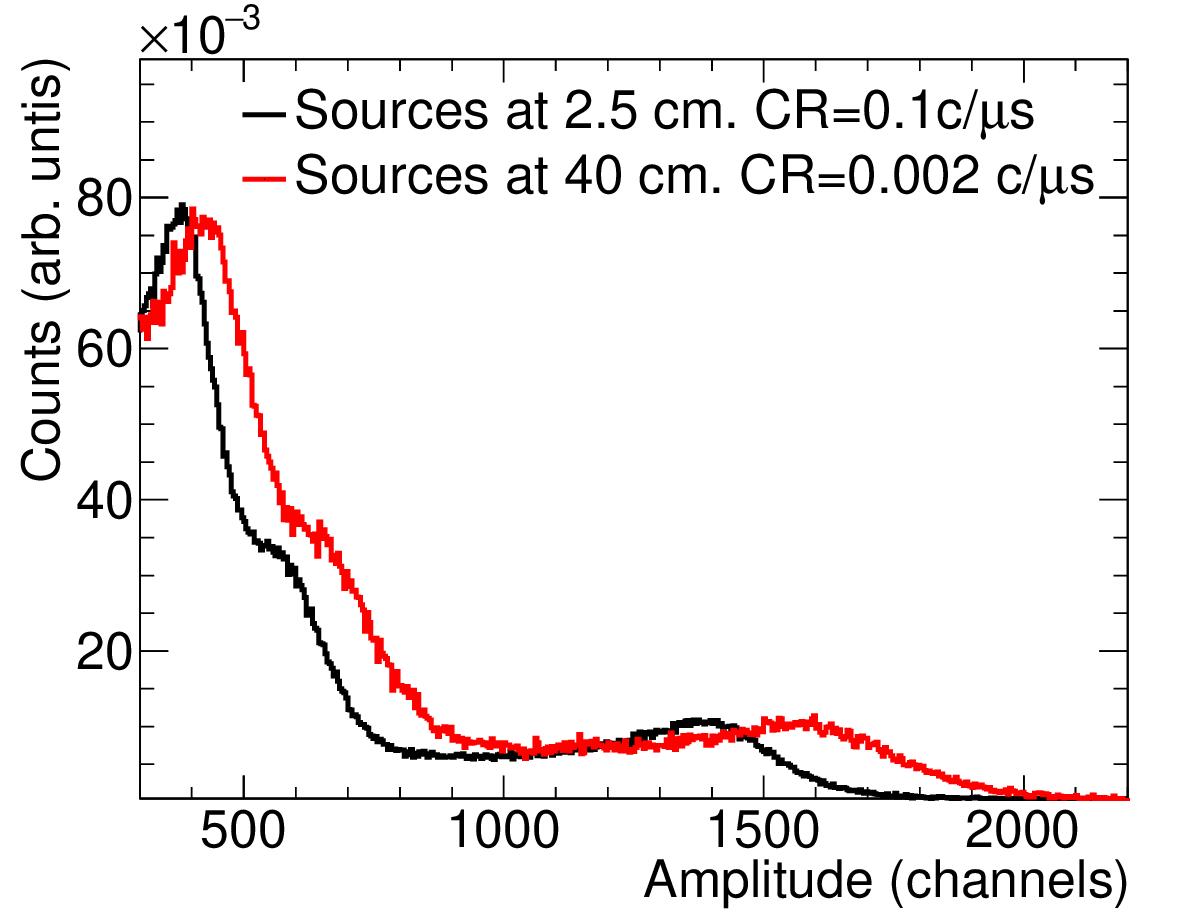}
\caption{Amplitude spectra for a combination of $^{137}$Cs and $^{88}$Y calibration sources with a total activity of $\sim$400 kBq placed at two different distances from a carbon-fiber housing detector. A gain shift of 12\% is observed from one measurement to the other. The Counting Rates (CR) obtained with an energy threshold of 0.15 MeV are also given in the figure.}
\label{fig:GainCRL6D6} 
\end{figure}

 \item Gain shifts due to the arrival of the particle flash of EAR2. The particle flash formed by relativistic charged particles, high-energy neutrons, and prompt $\gamma$-rays arriving at the experimental area at very short times ($<$1 $\mu$s) after the proton beam hits the spallation target induces a strong saturation of the detectors. As a result, gain shifts as a function of time of flight (i.e. neutron energy) appear during the recovery of the BICRON and carbon-fiber housing detectors. The gain of the detectors slowly recovers and after approximately 10 ms (corresponding to neutrons of approximately 0.02 eV) the gain is back to the same condition as before the particle flash. 
 This effect is illustrated in Fig. \ref{fig:GainEnergy}, where the deposited energy spectra for a $^{197}$Au sample are shown for different neutron energy ranges. The amplitude spectra should be very similar in the energy range of the figure. Although the visible difference (of spectra in Fig. \ref{fig:GainEnergy}) might be explained by gain change due to different absolute count rate, we have verified from a measurement of a radioactive $^{88}$Y source with beam that the gain change is produced by the particle flash \cite{Alcayne_22}. 
\begin{figure}[!ht]
\centering
\includegraphics[width=0.49\textwidth]{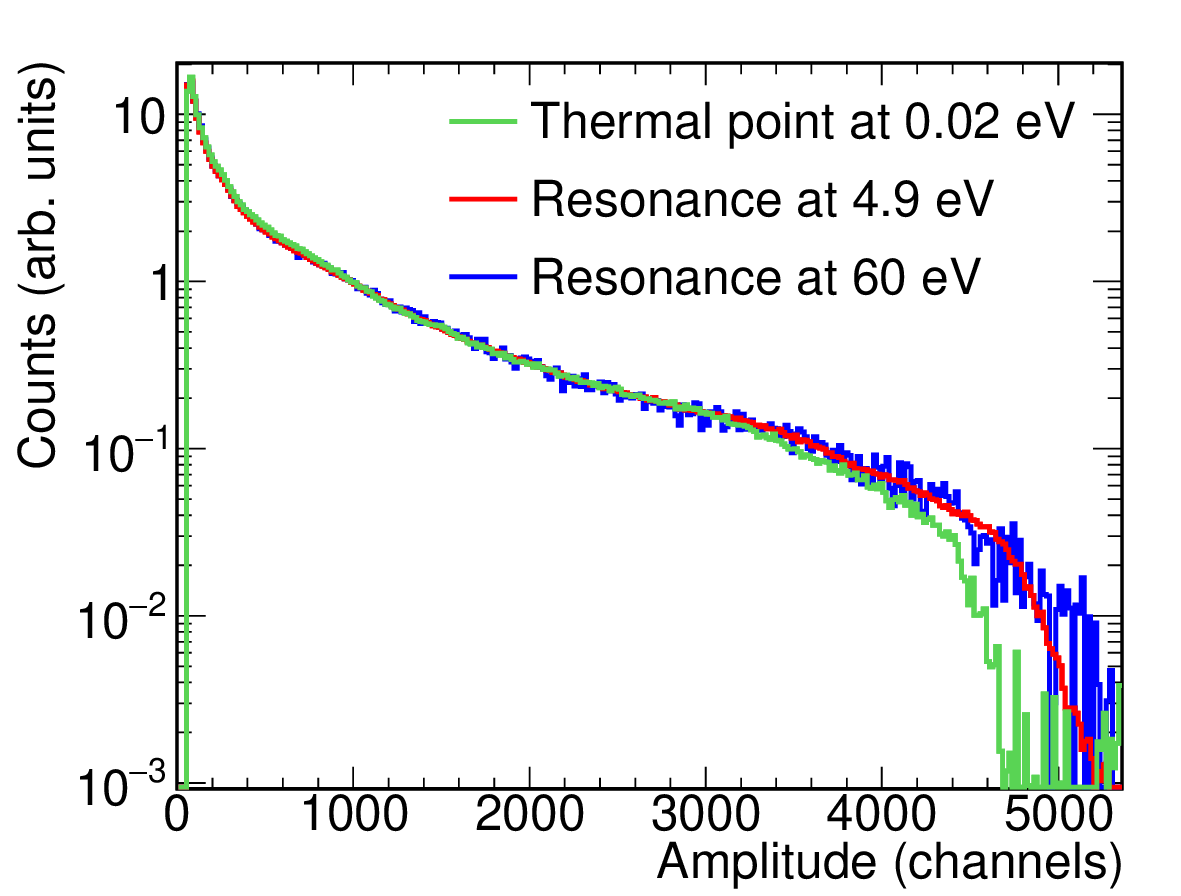}
\caption{Amplitude spectra obtained from a measurement with a BICRON detector at 5 cm from a $^{197}$Au sample of 0.5 cm diameter and 100 $\mu$m thickness. The neutron separation energy of $^{197}$Au is 6.512 MeV. The spectra (from different energy ranges) are normalized to the same number of detected counts. }
\label{fig:GainEnergy} 
\end{figure}

In Fig. \ref{fig:GainIntensities}, the deposited energy spectra for three different proton intensities in the same energy range are presented. It can be seen that the gain increases with higher proton intensity indicating that the effects of the particle flash increase with the beam intensities.

\begin{figure}[!ht]
\centering
\includegraphics[width=0.49\textwidth]{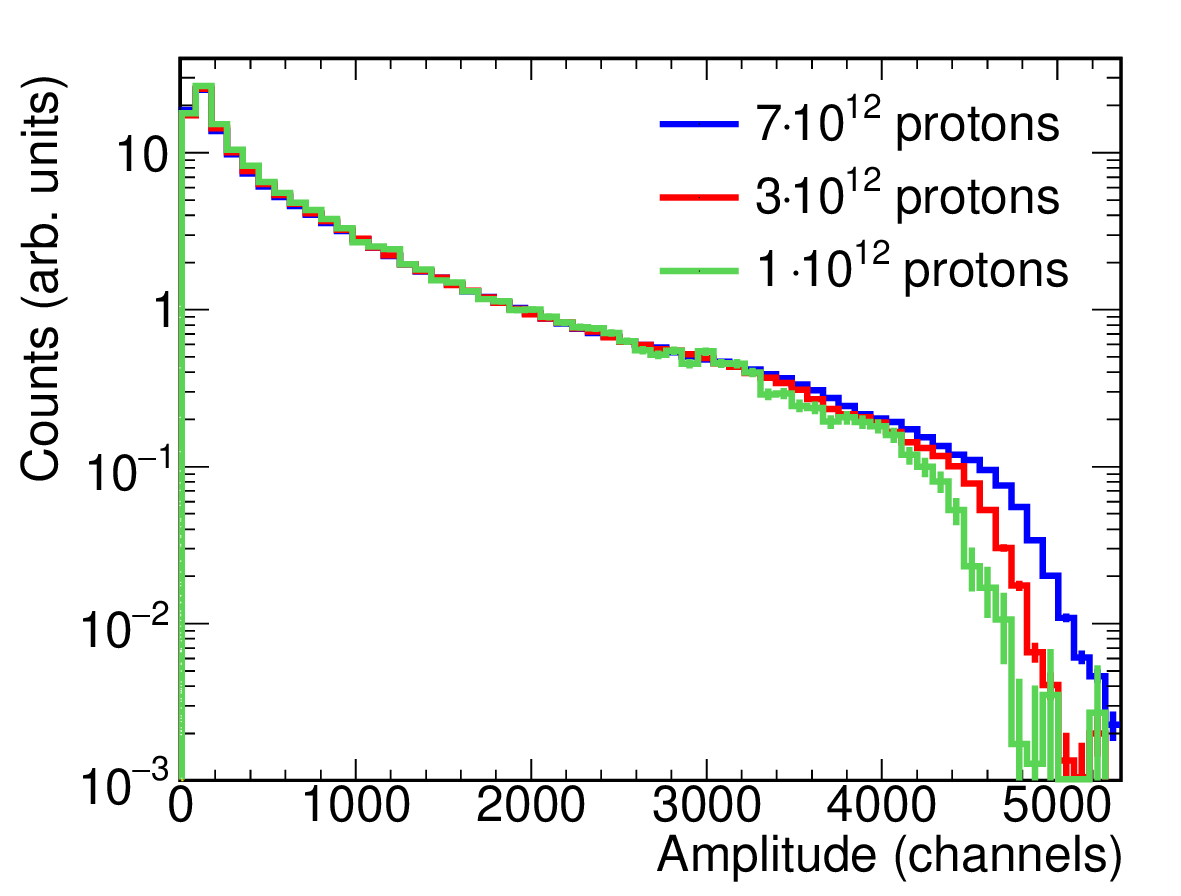}
\caption{Amplitude spectra obtained for the 4.9 eV resonance of a $^{197}$Au sample of 0.5 cm diameter and 100 $\mu$m thickness with a BICRON detector at 5 cm. Three different proton intensities are presented in the plot. The spectra (from different proton intensities) are normalized to the same number of detected counts.}
\label{fig:GainIntensities} 
\end{figure}

Note that the mass of the Au sample used in the tests presented in Figs. \ref{fig:GainEnergy} and \ref{fig:GainIntensities} is low enough to exclude pile-up effects.

\item Gain shifts produced by rapid counting rate variations as a function of the time of flight. It has been observed that the detectors show different gains at neutron energies above and below a strong resonance inducing a high counting rate. 
As shown in Fig. \ref{fig:GainCRL6D6Beam}, the gain of the detector at neutron energies 1.5-3.5 eV (below the strong 4.9 eV $^{197}$Au resonance) increases with respect to 6-8 eV (above the Au resonance), and then recovers at the thermal point. This effect shift goes in opposite direction than that expected due to the particle flash. Also, this effect cannot be due to pile-up due to the fact that the energy ranges are in the tails of the resonance in the region of relatively low cross section and also both energy ranges have been selected to have similar counting rates. Thus the effect is attributed to a gain shift produced by the high counting rate ($\sim$10 counts/$\mu$s) of the strong resonance at 4.9 eV.

 \begin{figure}[!ht]
\centering
\includegraphics[width=0.49\textwidth]{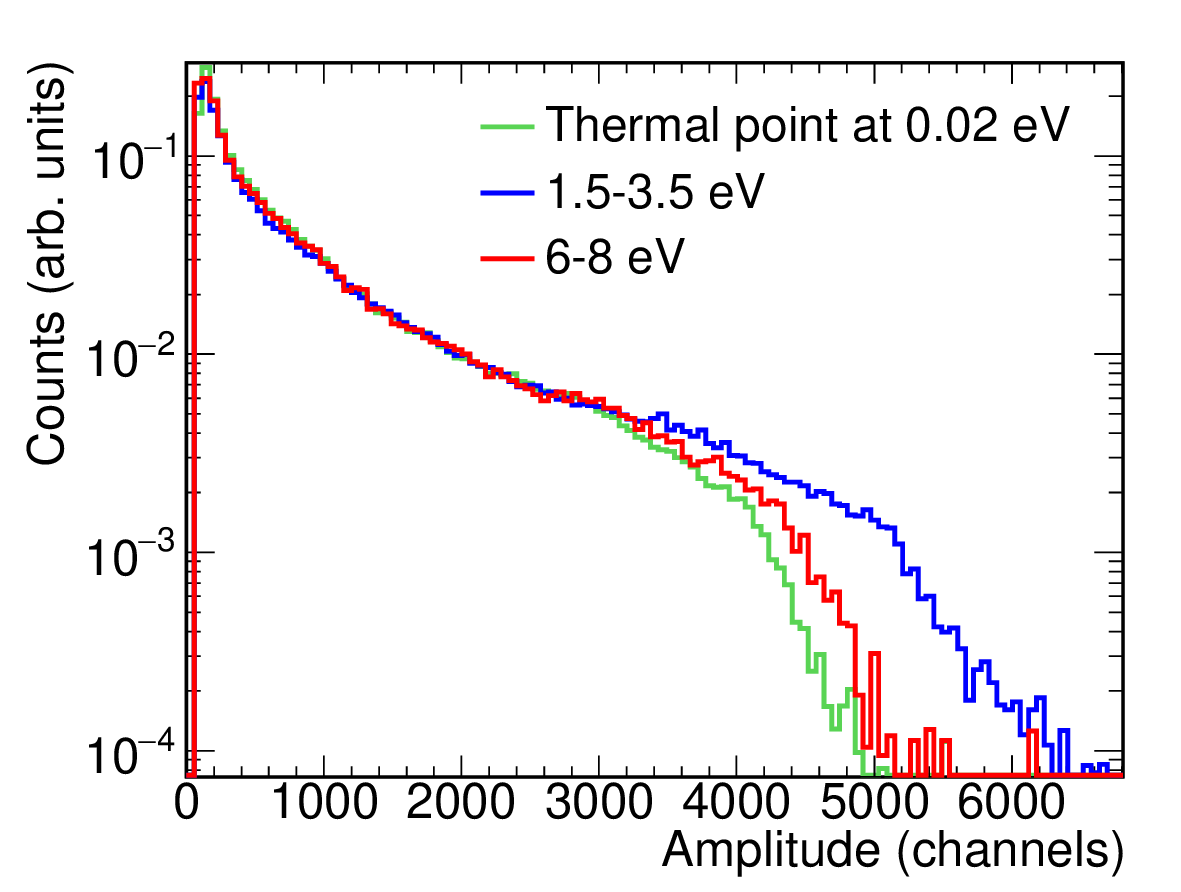}
\caption{Amplitude spectra obtained for measurements with a carbon-fiber housing detector at 10 cm from a $^{197}$Au sample of 2 cm diameter and 100 $\mu$m thickness. The spectra, normalized between them to the number of detected counts, are presented for different neutron energy ranges and the nominal proton intensity of 7$\cdot$10$^{12}$ protons per pulse.}
\label{fig:GainCRL6D6Beam} 
\end{figure}

 \end{itemize}

 \section{sTED description and specifications}\label{section:Detector}

The sTED has been specifically designed to improve the capture detection setup at EAR2, following the simple idea of reducing the counting rate per module by about one order of magnitude by replacing large volume C$_6$D$_6$ detectors with a much larger amount of smaller modules, forming an array with a comparable total efficiency. Each sTED module has an active volume of 0.044 liters, which is $\sim$14 times and $\sim$23 times smaller than the active volumes of the BICRON (0.621 liters) and the carbon-fiber housing (1.0 liters) detectors, respectively. In addition, smaller photomultipliers optimized for high counting rates are used to provide additional robustness. The following subsections present the technical specifications, the detector response to $\gamma$-rays, and the application of the PHWT to the data measured with an sTED array consisting of three modules.


 \subsection{Detector characteristics}\label{subsection:TecnicalSpecifications}
\begin{figure}[!ht]
\centering
\includegraphics[width=0.49\textwidth]{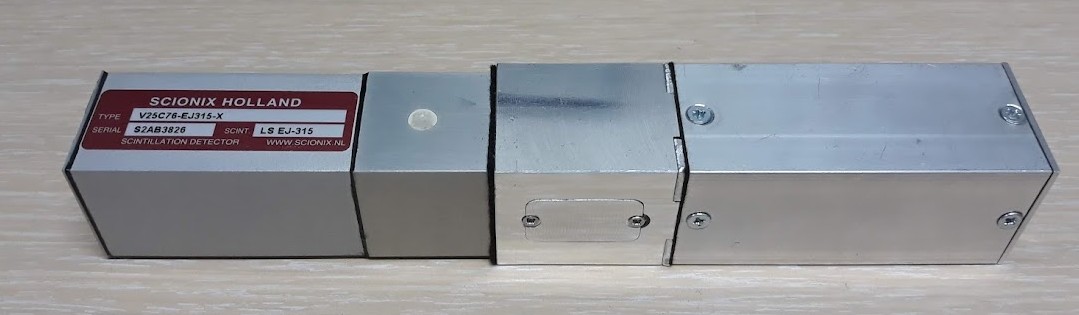}
\includegraphics[width=0.49\textwidth]{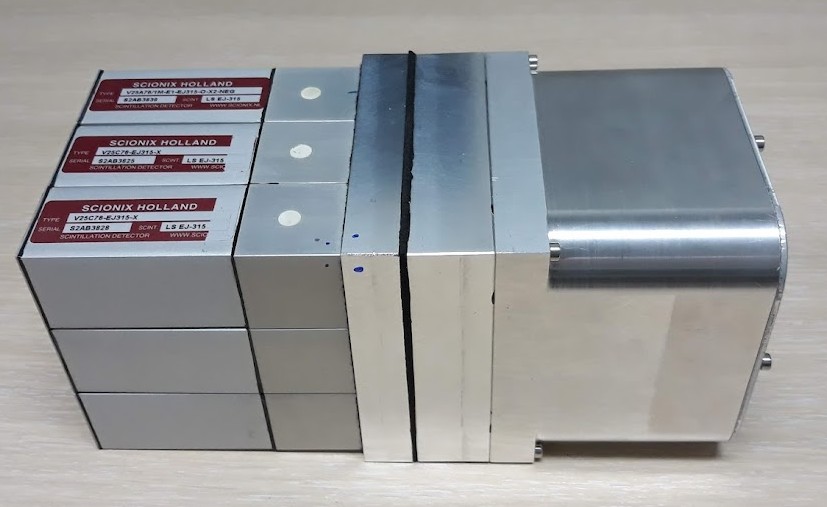}
\caption{Photos of one sTED module (top) and nine sTED modules grouped into a cluster (bottom).}
\label{fig:Picture_sTED} 
\end{figure}

\begin{figure}[!ht]
\centering
\includegraphics[width=0.49\textwidth]{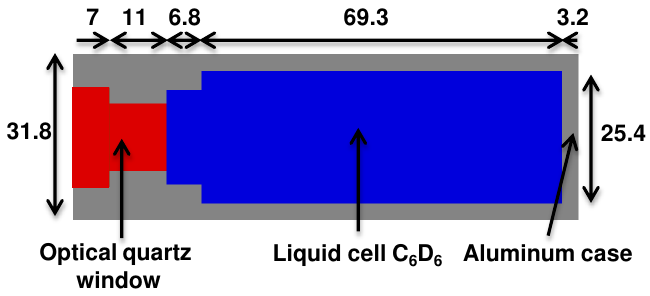}
\caption{Drawing of one sTED module with the different components and their sizes in millimeters. The PMT is coupled to the optical quartz window. }
\label{fig:DrawingSted} 
\end{figure}

The sTED modules were designed via Monte Carlo (MC) simulations and one prototype was tested in the laboratory at CIEMAT. Once validated, nine modules were purchased from Scionix \cite{Scionix}. Fig. \ref{fig:Picture_sTED} shows one module and a possible assembly of the nine modules. The C$_6$D$_6$ cell is coupled to the photomultiplier with an optical quartz window. The dimensions of the module and its elements are presented in Fig. \ref{fig:DrawingSted}.


Three different 1" Hamamatsu photo-multipliers \cite{Hamamatsu} coupled to an sTED module were tested: R5611A, R2076 and R11265U-100. It was found that R5611A and R2076 also suffered from gain shifts when exposed to high counting rates of $\sim$0.05 counts/$\mu$s, whereas the R11265U-100 showed a stable gain up to at least 0.25 counts/$\mu$s. This photomultiplier has a borosilicate window and Super Bialkali (SBA) photocathodes with decoupling capacitors and a tapered voltage divider distribution ratio specially designed for high counting rates.

\begin{figure}[!ht]
\centering
\includegraphics[width=0.49\textwidth]{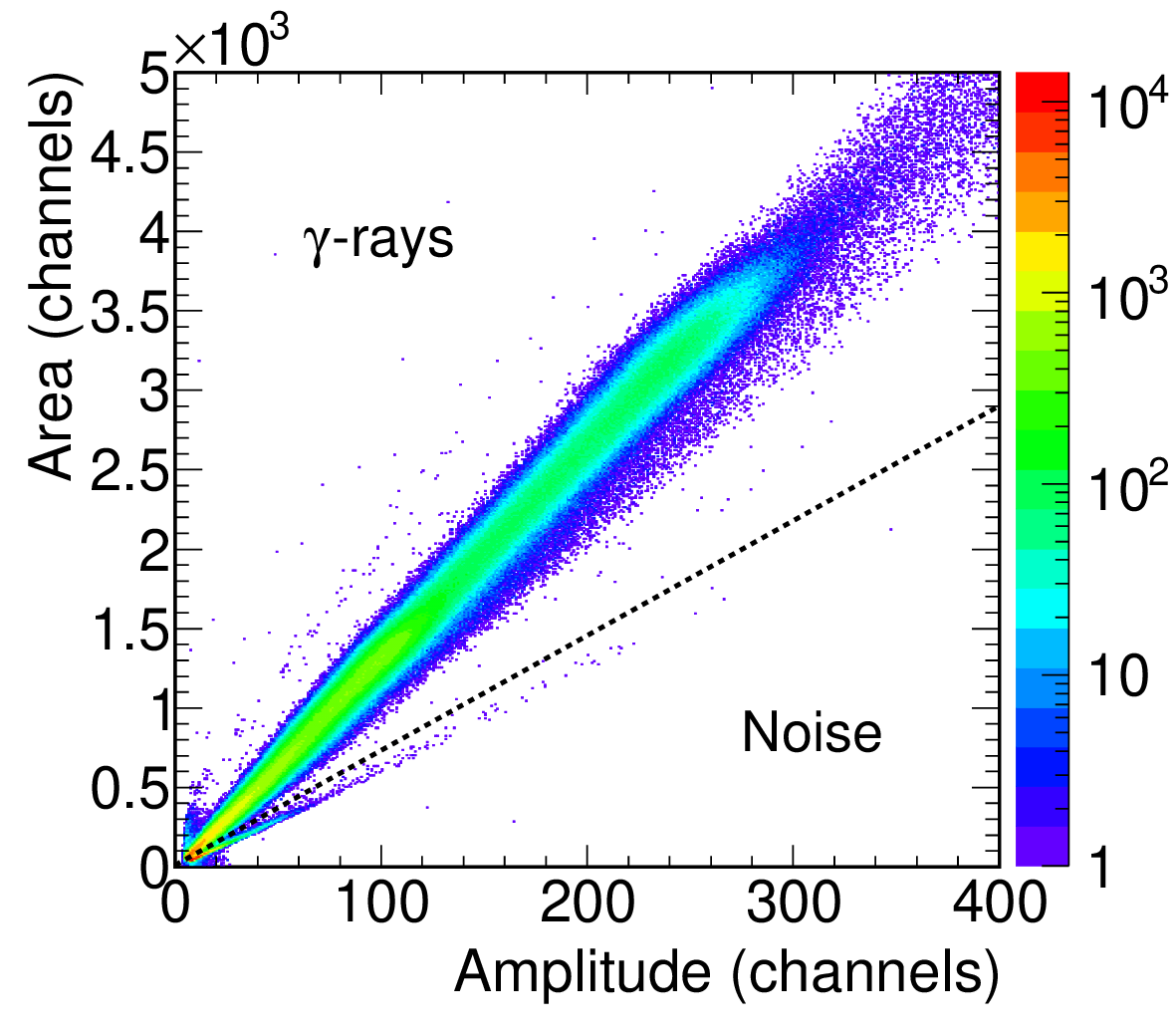}
\includegraphics[width=0.49\textwidth]{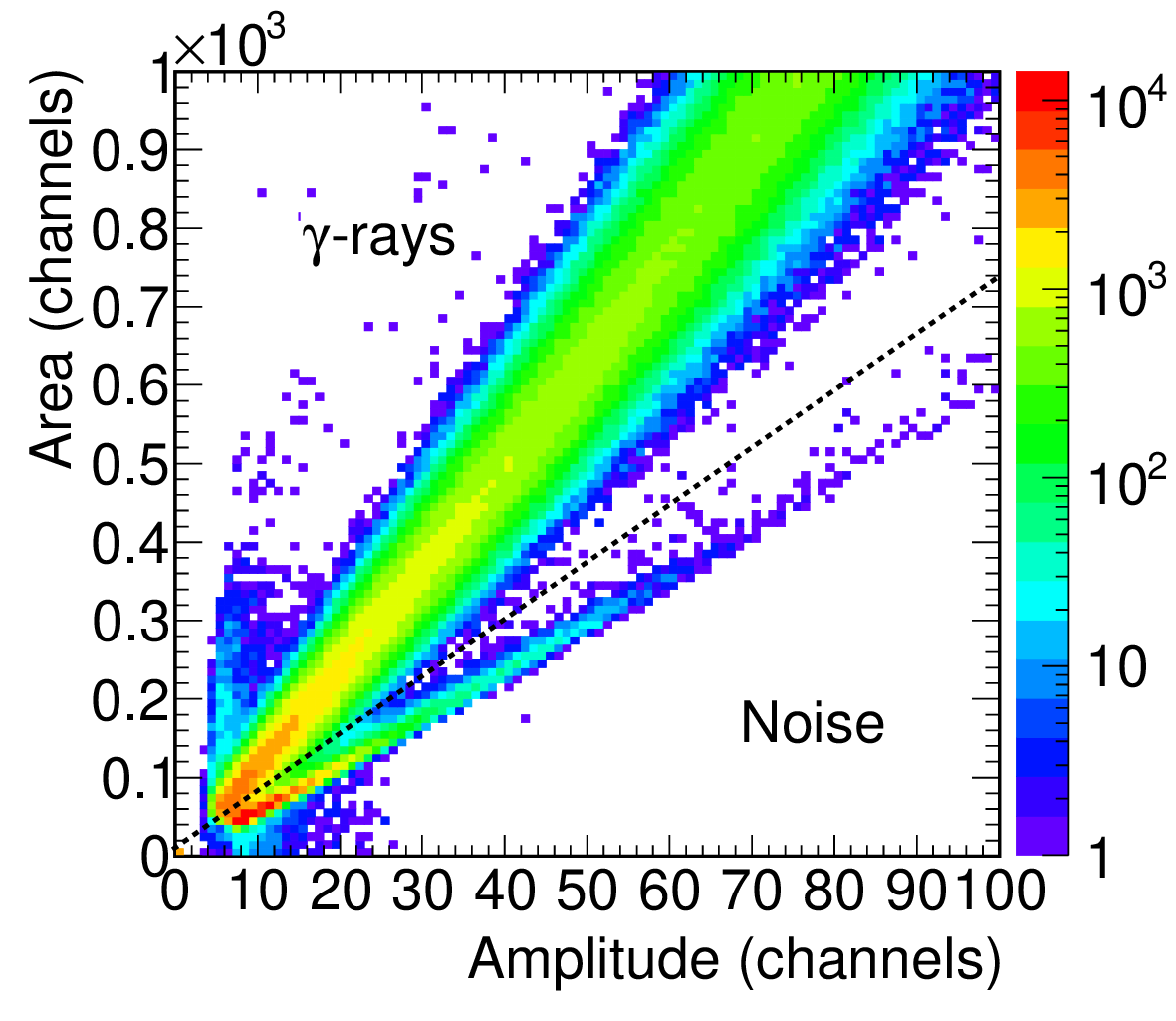}
\caption{2D plot showing the amplitude versus the area of the sTED signals with an R11265U-100 photomultiplier when measuring an $^{88}$Y calibration source. The black dashed line separates two different types of signals, see text for details. The bottom figure is a zoom of the top one.}
\label{fig:2D_Area_Amp} 
\end{figure}

 In addition, we have performed a detailed study of the shape of the sTED signals. As can be seen in Fig. \ref{fig:2D_Area_Amp}, there are two types of signals. The ones with a high area-to-amplitude (above the dotted line) are due to $\gamma$-rays depositing energy in the scintillation liquid. The remaining signals have a different origin, which was attributed to the instantaneous production of one or a few photo-electrons in the photo-cathode of the photo-multiplier \cite{Knoll_79}. These signals are called \textit{noise} in this work because they are not produced by $\gamma$-rays interacting with the C$_6$D$_6$ liquid.

\begin{figure}[!ht] 
\centering
\includegraphics[width=0.49\textwidth]{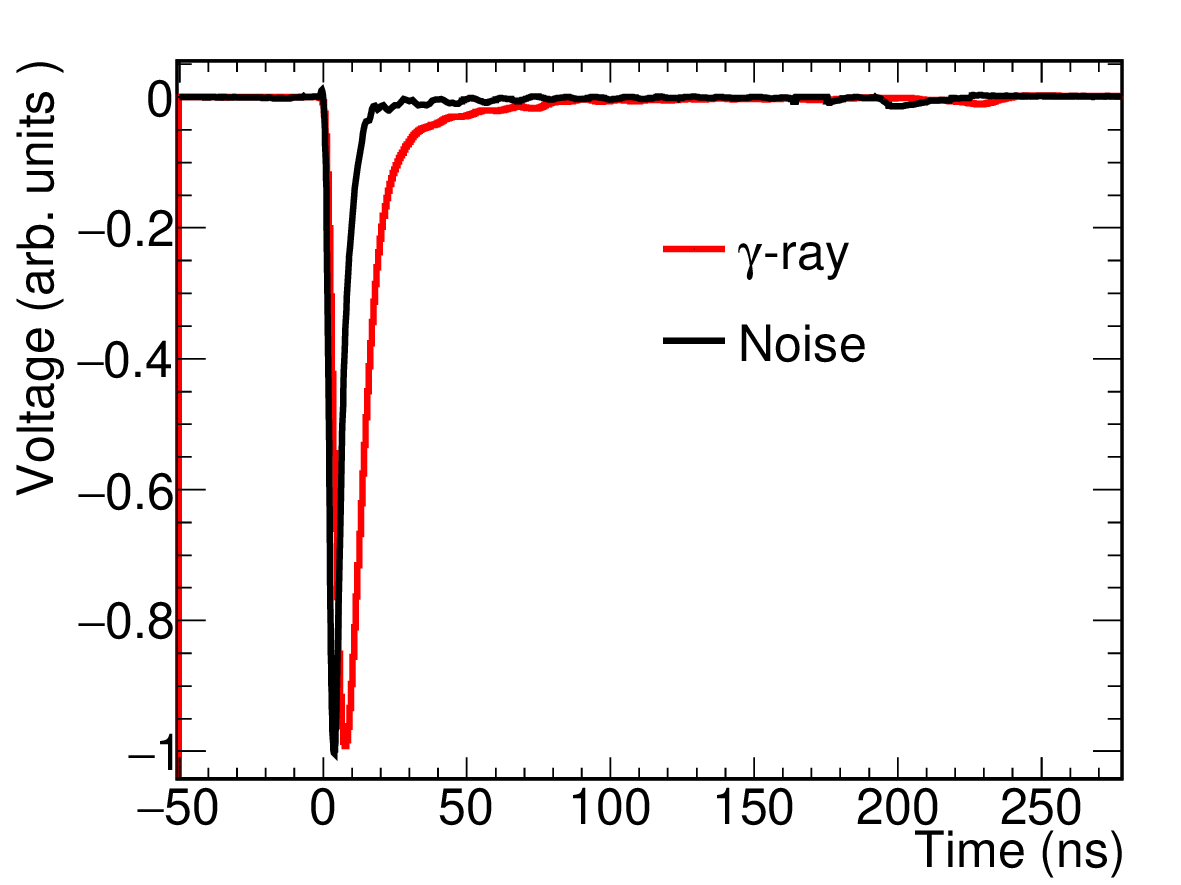}
\caption{Average sTED signals produced by $\gamma$-rays and \textit{noise}. The signals are normalized to the same maximum amplitude.}
\label{fig:AvgSignal} 
\end{figure}

At n\_TOF, the signals are digitized, stored, and then analyzed using dedicated pulse shape analysis routines \cite{Zugec_16}. Pulse shape fitting is found to be an optimal technique to discriminate the two types of signals \cite{Guerrero_08}. To perform the discrimination two average signal shapes (depicted in Fig. \ref{fig:AvgSignal}) have been obtained, one for signals related to $\gamma$-rays and the other to the \textit{noise}. The use of the average signal shapes in the fitting also allows to mitigate pile-up effects

\begin{figure}[!ht]
\centering
\includegraphics[width=0.49\textwidth]{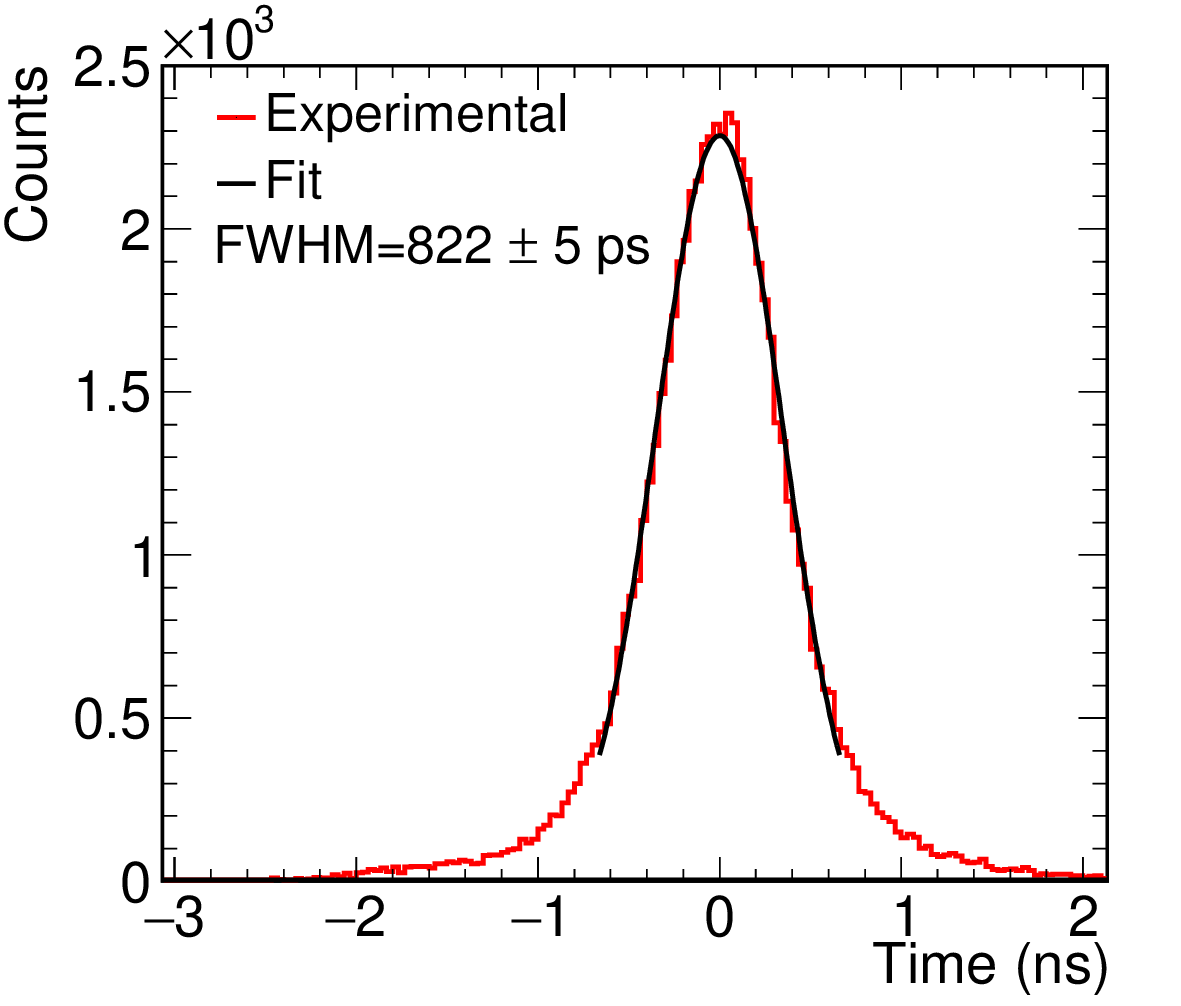}
\caption{Distribution of the time differences between signals in coincidence between an sTED module and a LaBr$_3$ detector, when measuring a $^{60}$Co source. The FWHM of the distribution is 822 $\pm$ 5 ps, including the contribution of the LaBr$_3$ and the sTED, see text for details.}
\label{fig:TimeResol} 
\end{figure}

 The time resolution of one sTED module was measured with respect to a LaBr$_3$ detector with 354 $\pm$ 4 (statistical) $\pm$ 10 (systematic) ps time resolution \cite{Gramage_16}. The coincident signals corresponding to the detection of the 1173 and 1332 keV $\gamma$-rays from a $^{60}$Co source in each detector were used to determine the time resolution, assuming that both detectors have a Gaussian time response. The distribution of the time differences between the coincident sTED and LaBr$_3$ events is shown in Fig. \ref{fig:TimeResol}. The obtained time resolution of an sTED module is 742 $\pm$ 15 ps, including both statistical and systematic uncertainties.
 
 One should notice that the reduction of the active volume and thus the need of using a thick solid housing for keeping the liquid scintillator and the use of a PMT with a borosilicate glass window (instead of quartz) could have the drawback of increasing the neutron sensitivity \cite{Plag_03} compared to larger volume detectors with similar housing thicknesses. The neutron sensitivity of the sTED was estimated by Monte Carlo simulations for capture measurements of various isotopes. 
 
 The sensitivity values are given in Table \ref{table:nsenstivity}, where the magnitude of the background due to scattered neutrons is estimated for some representative resonance in certain nuclei. There, $E_{n}$ is the resonance energy; $\Gamma_n$ and $\Gamma_\gamma$ are its neutron and radiative widths, respectively; $\varepsilon_n$ is the probability of detecting a neutron scattered in the sample with energy $E_{n}$; $\varepsilon_\gamma$ is the efficiency of detecting the corresponding (n,$\gamma$) cascade, and $(\varepsilon_n/\varepsilon_\gamma)\cdot(\Gamma_n/\Gamma_\gamma)$ estimates the size of the background due to elastically scattered neutrons compared to the (n,$\gamma$) detected events.
\begin{table}[!ht]
\caption{Estimation of the neutron sensitivities ($(\varepsilon_n/\varepsilon_\gamma)\cdot(\Gamma_n/\Gamma_{\gamma})$) of one sTED module for different nuclei and resonances. For details see the text.}
\centering
\begin{tabular}{ccccc}
\hline
Isotope &E$_n$ (eV)& $\frac{\Gamma_n}{\Gamma_{\gamma}}$ & $\frac{\varepsilon_n}{\varepsilon_\gamma}$ & $(\varepsilon_n/\varepsilon_\gamma)\cdot (\Gamma_n/\Gamma_\gamma)$\\ 
 \hline
$^{197}$Au &4.91 & 1.2$\cdot 10^{-1}$ & 1.6$\cdot 10^{-3}$ &2.0$\cdot 10^{-4}$\\
$^{240}$Pu & 5.01 & 8.4$\cdot 10^{-2}$ & 1.6$\cdot 10^{-3}$& 1.4$\cdot 10^{-4}$ \\
$^{244}$Cm & 7.66 &	4.9 & 1.6$\cdot 10^{-3}$ & 8.0$\cdot 10^{-3}$ \\
$^{244}$Cm & 86.1 & 6.6$\cdot 10^{-1}$ & 5.5$\cdot 10^{-4}$ & 3.6$\cdot 10^{-4}$ \\
$^{207}$Bi &12100 &2.2$\cdot 10^{3}$&1.1$\cdot 10^{-4}$&2.4$\cdot 10^{-1}$\\
$^{207}$Pb &41100 &3.7$\cdot 10^{2}$&2.3$\cdot 10^{-4}$&8.4$\cdot 10^{-2}$\\

\hline
 \end{tabular}

\label{table:nsenstivity}
\end{table}

The values $(\varepsilon_n/\varepsilon_\gamma)\cdot(\Gamma_n/\Gamma_\gamma)$ given in Table~\ref{table:nsenstivity} indicate that the neutron scattering background in the resonances of $^{240}$Pu, $^{244}$Cm and $^{197}$Au are $\lesssim$ 0.1\%. However, for resonances with unfavorable $(\Gamma_n/\Gamma_\gamma)$ values like the ones in $^{207}$Pb and $^{209}$Bi, the neutron-induced background would be as large as 8.4\% and 24\%. In general, the increase of the ratio of $\Gamma_n/\Gamma_\gamma$ with neutron energy will require corrections of a few percent for almost all the nuclei at energies above 10 keV. For measurements targeting on these cases, highly optimized C$_6$D$_6$ detectors such as the ones with a carbon-fiber housing, thinner optical (or even no at all) windows, and PMTs with quartz windows \cite{Plag_03} could be required.

 \subsection{Detector response to $\gamma$-rays}\label{section:DetectorResponse}

\begin{figure}[!ht]
\centering
\includegraphics[width=0.49\textwidth]{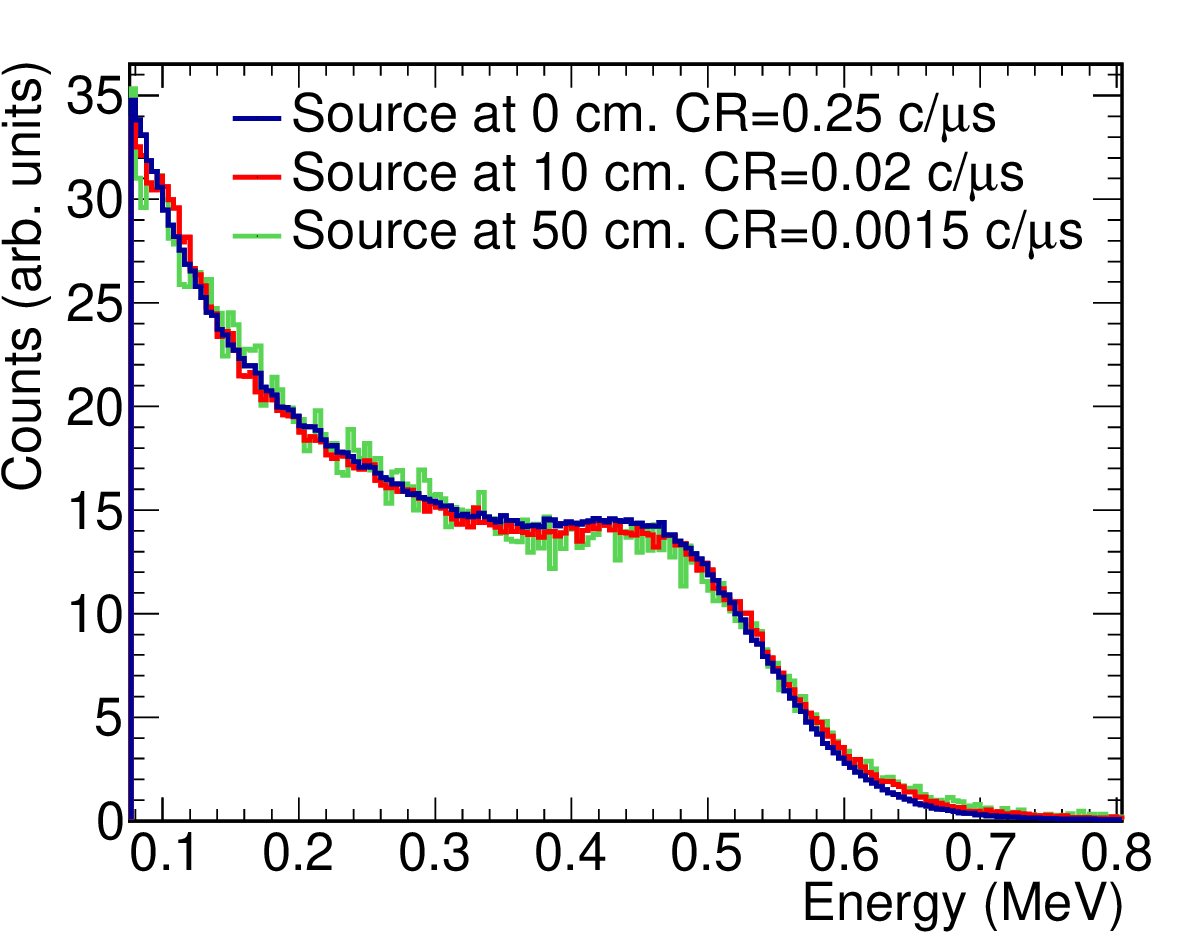}
\caption{Deposited energy spectra for a 20 MBq $^{137}$Cs calibration source placed at different distances from an sTED module with an R11265U-100 photomultiplier, along with the Counting Rates (CR) obtained using a deposited energy threshold of 0.15 MeV.}
\label{fig:GainCR} 
\end{figure}

As described in section \ref{section:Limitations}, the BICRON and carbon-fiber housing detectors exhibited gain shift when exposed to high activity $\gamma$-ray calibration sources. The response of the sTED modules with the R11265U-100 photomultiplier was investigated with a 20 MBq $^{137}$Cs source placed at different distances from the detector. The comparison of the $^{137}$Cs pulse height spectra recorded at 0.25, 2$\cdot10^{-2}$ and 1.5$\cdot10^{-3}$ c/$\mu$s are shown in Fig. \ref{fig:GainCR}. It can be concluded that for counting rates as large as 0.25 c/$\mu$s no gain shifts are observed.
 
The linearity and the energy resolution of the detector were characterized with six $\gamma$-ray sources: $^{133}$Ba, $^{137}$Cs, $^{207}$Bi, $^{60}$Co, $^{88}$Y and AmBe using the Compton edge clearly visible in the spectra. There are almost no signals corresponding to the full-energy peak in the sTED detector with these $\gamma$-ray sources. The procedure used consisted of:
\begin{enumerate}[label=(\roman*)]
 \item Gaussian folding of the simulated Monte Carlo spectra of deposited energies from the $\gamma$-ray calibration source.
 \item Energy re-calibration of the experimental data.
 \item Repetition of points i) and ii), until the best fit of spectra near the Compton edge is reached
 \item In the region near the Compton edge, the relation between the amplitude of the signals and the deposited energy in the detectors ($E$) have been determined for each Compton edge. Also, the resolution ($\Delta E/E$) in these regions have been determined.

 \item Least square fit of to the values obtained in the point iv) to linear and parabolic energy calibrations for determining the energy calibration curve.
 \item Fit of the $\Delta E/E$ values to the $\mathrm{\Delta E/E=2.35\cdot \sqrt{\alpha/ E+\beta}}$ resolution function.
\end{enumerate}

In order to perform the sTED calibration with the described procedure, Monte Carlo simulations have been performed with a detailed geometric description of the sTED modules and the well-validated Standard Electromagnetic physics package of Geant4 \cite{Agostinelli_03}.

\begin{figure}[!ht]
\centering
\includegraphics[width=0.49\textwidth]{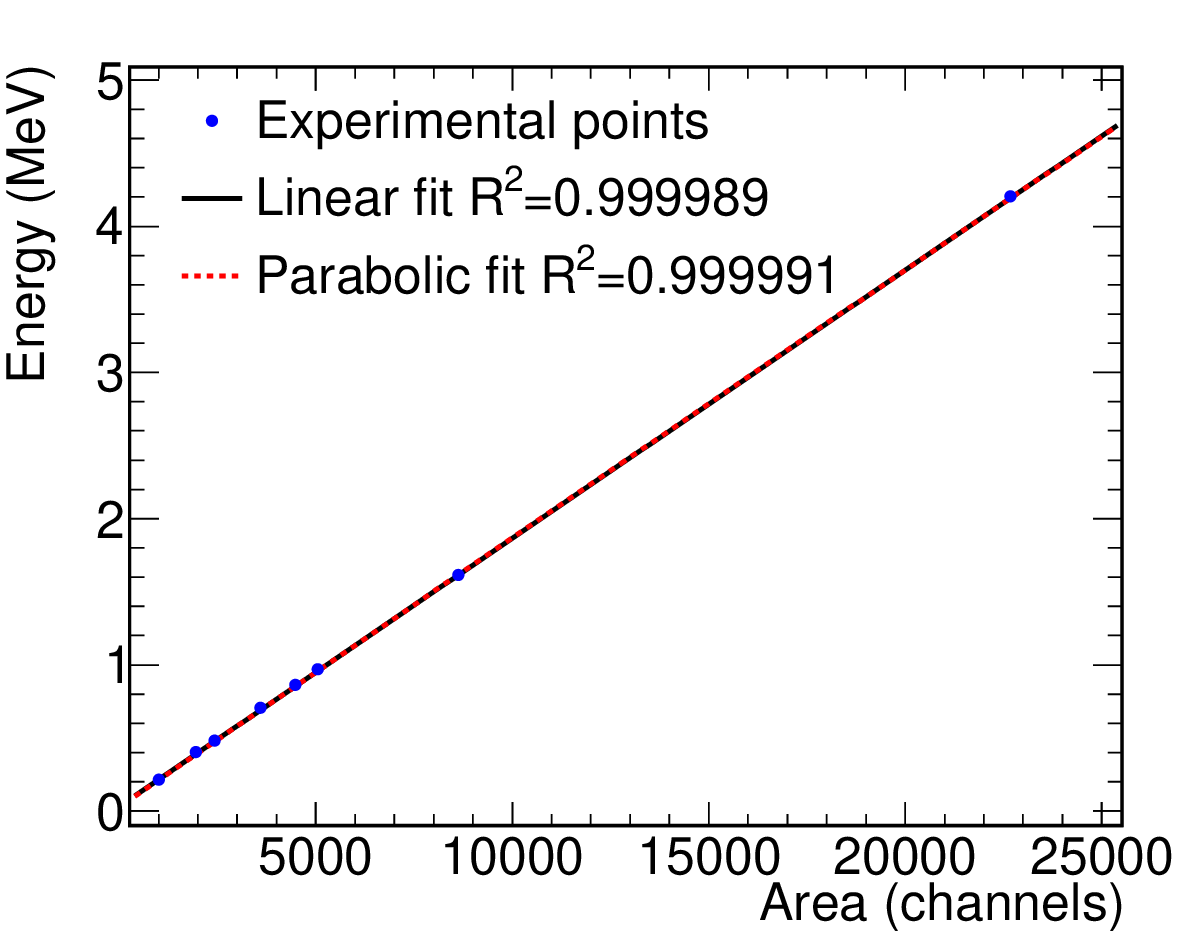}
\includegraphics[width=0.49\textwidth]{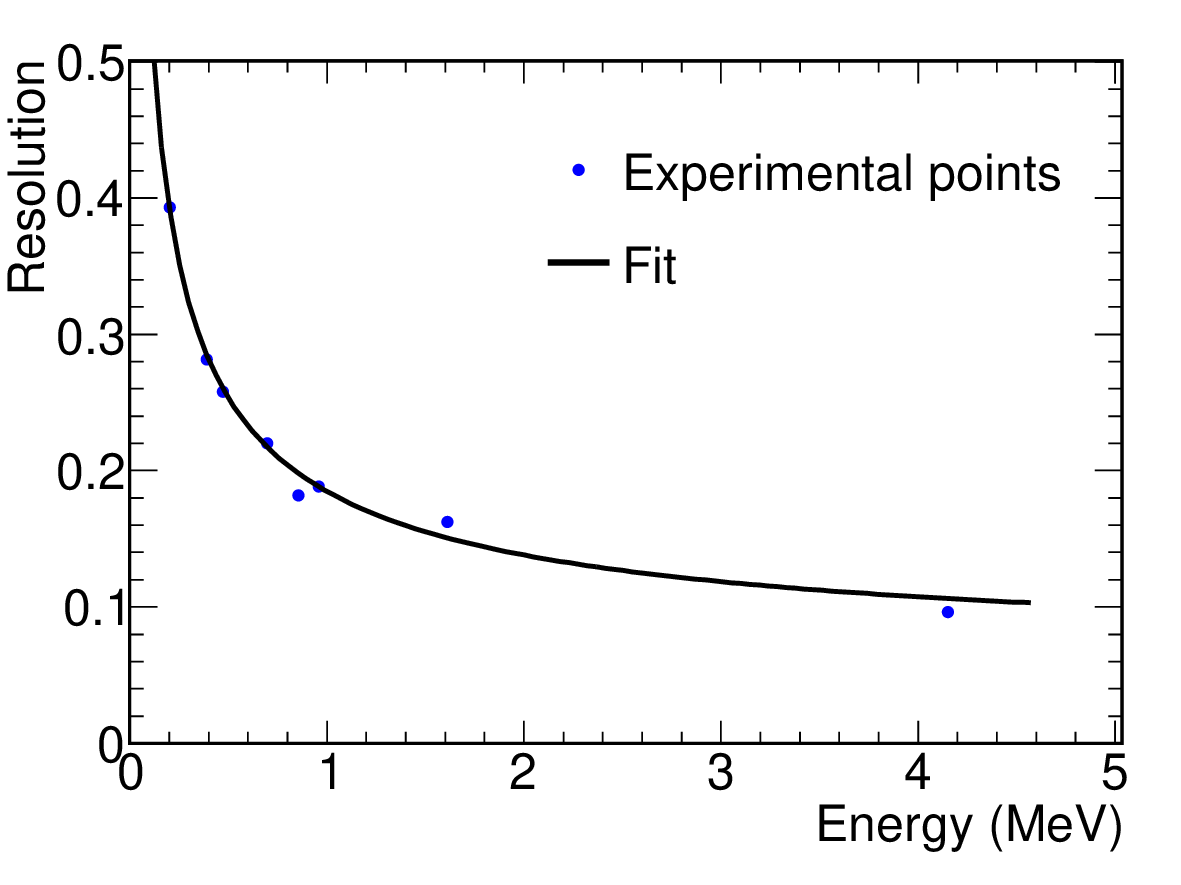}

\caption{Top panel: deposited energy in the detector as a function of the area of the signals (blue points corresponding to Compton edges). The values have been fitted to a straight line (black dashed line) and a parabola (red dashed line). Bottom panel: energy resolution ($\Delta E/E$) of one sTED module as a function of the deposited energy (blue points). The experimental points have been fitted with the function: $\mathrm{\Delta E/E=2.35\cdot \sqrt{\alpha/E+\beta}}$. The results of the fit are $\alpha=0.00545$ in MeV and $\beta=0.000729$.}
\label{fig:Calibration} 
\end{figure}

\begin{figure}[!ht]
\centering
\includegraphics[width=0.49\textwidth]{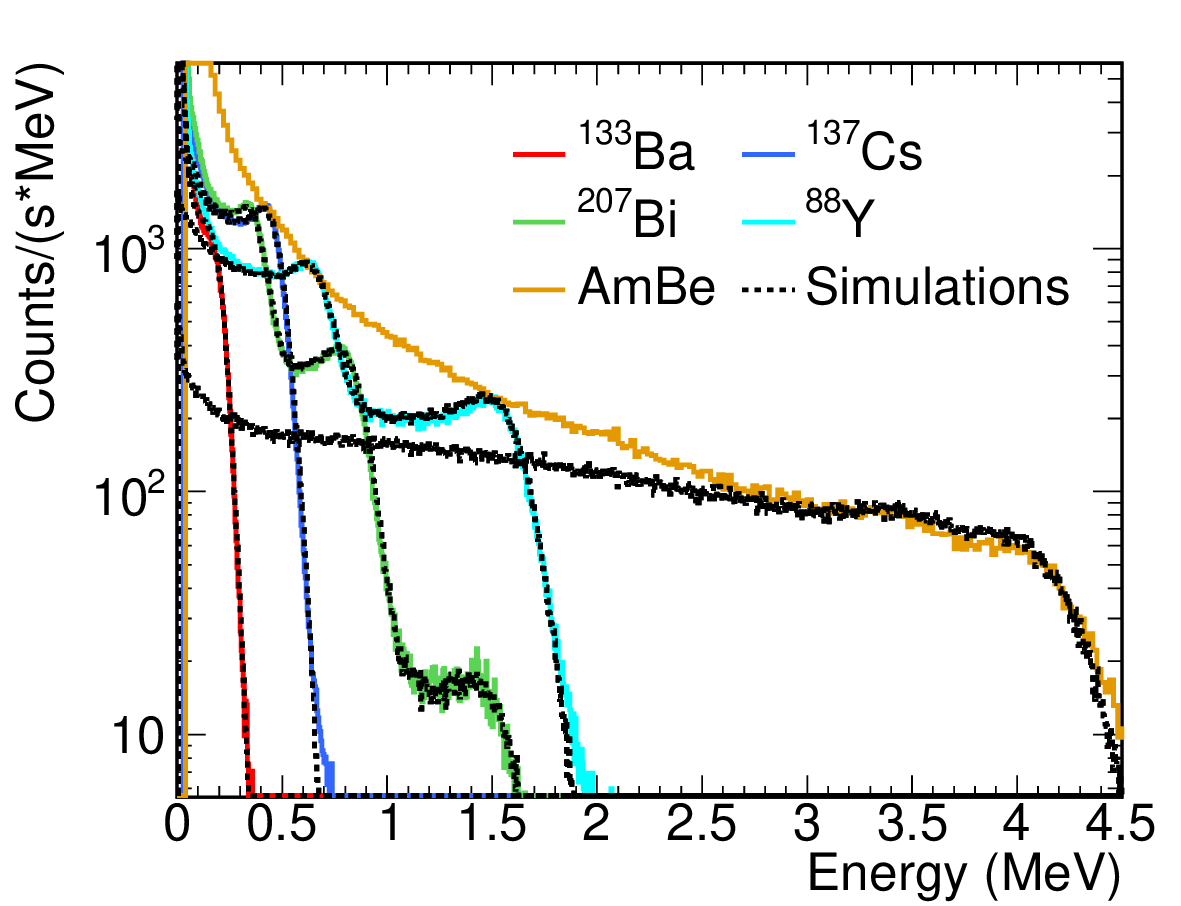}
\caption{Experimental deposited energy spectra obtained with an sTED module for various $\gamma$-ray sources ($^{133}$Ba, $^{137}$Cs, $^{207}$Bi, $^{88}$Y and AmBe) compared with Geant4 simulations. The area of the experimental spectra are normalized to the simulated ones.}
\label{fig:ResultsCalibration} 
\end{figure}

The top panel of Fig. \ref{fig:Calibration} shows the least square fits of linear and parabolic functions to the experimental Compton edges for the different $\gamma$-ray energies. It can be seen from the coefficients of determination ($R^2$) that the linear and parabolic curves reproduce with equal accuracy the data and hence the calibration can be assumed to be linear. The fit of the detector energy resolution is presented in the bottom panel of Fig. \ref{fig:Calibration}. The resolution function obtained provides resolution values for this detector of 18\% at 1 MeV and 10\% at 5 MeV. The excellent agreement between the experimental and simulated spectra for the six $\gamma$-ray sources are shown in Fig. \ref{fig:ResultsCalibration}, indicating the high quality of the energy calibration and energy resolution determination. The differences for the AmBe spectra below 3 MeV are related to the response to the neutrons also emitted by the source, which were not simulated.

\begin{figure}[!ht]
\centering
\includegraphics[width=0.49\textwidth]{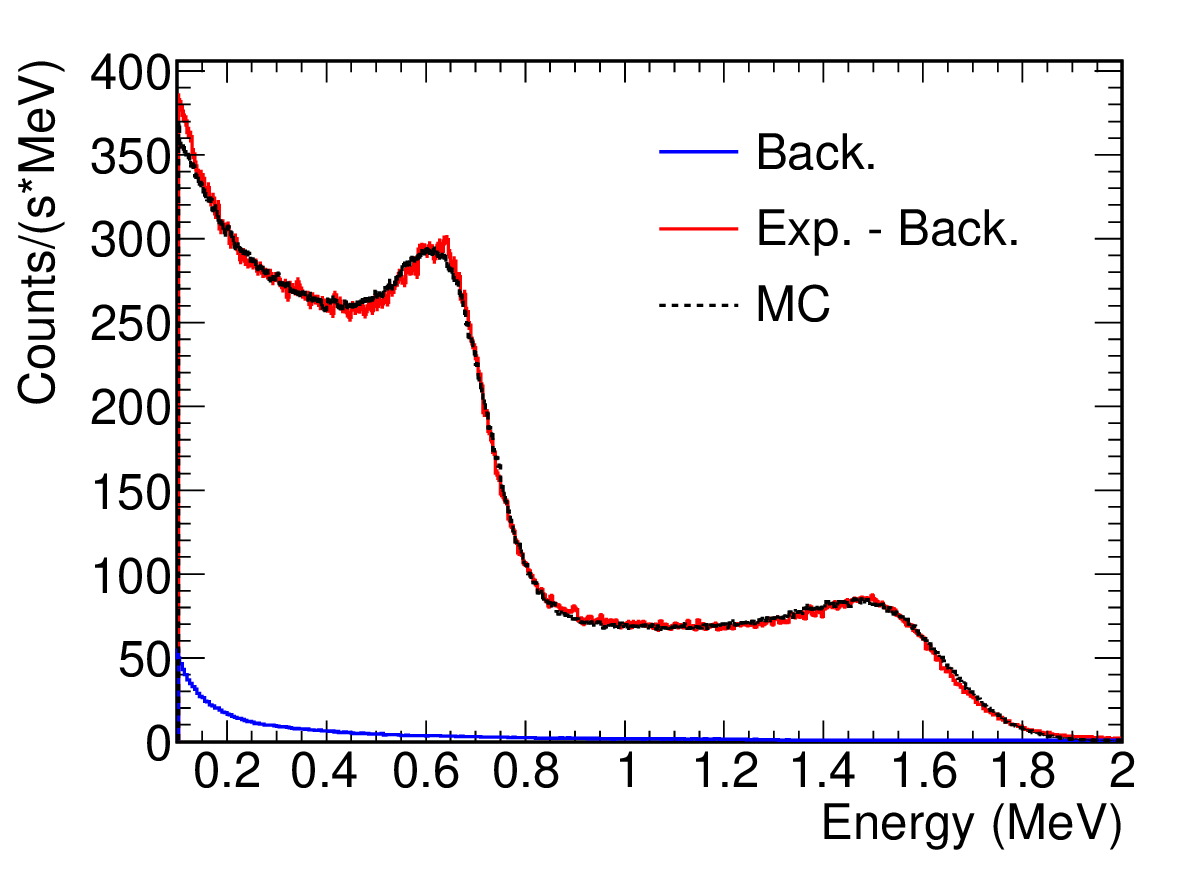}
\caption{Deposited experimental energy spectra (Exp.-Back.) after subtracting the background (Back.) for an $^{88}$Y $\gamma$-ray sources of 44.3 $\pm$ 1.3 kBq compared with Monte Carlo simulations (MC). The MC simulations have been scaled by 1.0027 to normalize to the experimental results.}
\label{fig:Y88} 
\end{figure}

It is important to notice that the Monte Carlo simulations are also used in the experimental technique applied in the analysis of (n,$\gamma$) cross-section measurements, described in detail in Section \ref{section:MeasuringTechnique}. For this reason, the overall quality of the simulations has been assessed with an absolute measurement of a well-characterized $^{88}$Y source (44.3 $\pm$ 1.3 kBq) at 5.0 $\pm$ 0.1 cm from an sTED module. As it can be seen in Fig. \ref{fig:Y88}, the Geant4 simulations folded with the energy resolution reproduce very accurately the experimental response. The small difference of 2.7\% between experiment and Geant4 based simulations is compatible with the uncertainty of the activity of the $^{88}$Y calibration source. These validated Geant4 simulations have been used to determine the absolute efficiency to detect various $\gamma$-ray decays. The simulated efficiencies are presented in Table \ref{table:eff} for the sTED, the BICRON, and the carbon-fiber housing detectors placed at 5 cm from the $\gamma$-ray emission point.

\begin{table}[!ht]
 \caption{Detection efficiencies, expressed as percentages, for three detectors. The efficiencies are for the detection of the $\gamma$-ray decay of $^{137}$Cs and $^{88}$Y, as well as for the (n,$\gamma$) cascades of $^{197}$Au and $^{240}$Pu computed with the NuDEX code \cite{Mendoza_23,Mendoza_20}. The efficiencies were calculated with Monte Carlo simulations by simulating the detectors at 5 cm from the $\gamma$-ray emission point and for a deposited energy threshold of 0.15 MeV. Note that for each decay/cascade, the efficiency scales approximately with the volume of the detector.}
\centering
\begin{tabular}{ccccc}
\hline
& $^{137}$Cs&$^{88}$Y&$^{197}$Au(n,$\gamma$)&$^{240}$Pu(n,$\gamma$)\\ 
& decay&decay&cascade&cascade\\ \hline
Carbon-fiber& 2.87 &6.15 &4.77 &6.09\\
BICRON & 2.28 &4.96 &3.86 &4.89 \\
sTED module & 0.20 &0.46 &0.35 &0.43 \\ \hline
 \end{tabular}

\label{table:eff}
\end{table}


 \subsection{The pulse height weighting technique}\label{section:MeasuringTechnique}
 The sTED has been designed to measure capture cross-sections using the Pulse Height Weighting Technique (PHWT) \cite{Macklin_67,Tain_04}, in which the efficiency of the detection system is transformed to become proportional to the total energy of the (n,$\gamma$) cascade, and therefore independent of the de-excitation pattern. The main conditions to be fulfilled for the applicability of this technique are: the $\gamma$-ray detection efficiency ($\varepsilon_\gamma$) is low, i.e. $\varepsilon_\gamma \ll 1$, and proportional to the $\gamma$-ray energy (E$_\gamma$), i.e. $\varepsilon_\gamma (E_\gamma ) = k \cdot E_\gamma$. For many detectors, such as the sTED, the efficiency to detect a $\gamma$-ray is not proportional to its energy. In these cases, ‘‘a posteriori’’ manipulation of the detector response function can be applied to make the detector response proportional to the energy of the $\gamma$-rays \cite{Macklin_67,Tain_04}. This is done by applying a weight to each recorded count dependent on its energy (pulse height), determined by the so-called Weighting Function (WF).

For the case of an array of many sTED modules, the total $\gamma$-ray detection efficiency of the setup can increase considerably. However, the PHWT can be still used to obtain the capture cross-section as presented in reference \cite{Mendoza_23}, as long as the intrinsic efficiency of each module remains small ($\varepsilon_\gamma \ll 1$). For the case of one sTED module, the efficiency to detect $\gamma$-rays from the (n,$\gamma$) cascades that are typically in the order of a few MeVs is low enough due to its small active volume, as seen in Table \ref{table:eff}.

\begin{figure}[!ht]
\centering
\includegraphics[width=0.49\textwidth]{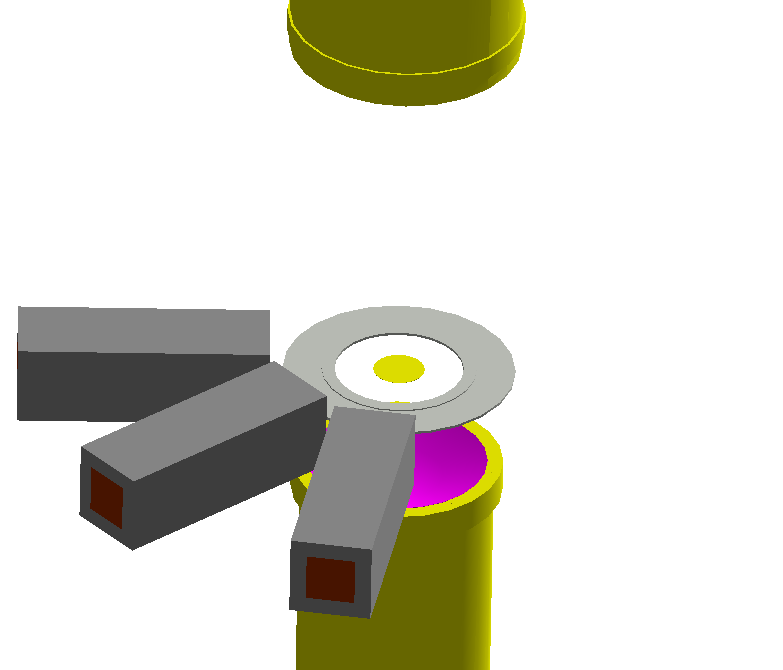}
\caption{A schematic view of of the setup simulated in Geant4 with three sTED modules and the gold sample.}
\label{fig:Geant4Setup} 
\end{figure}

\begin{figure}[!ht]
\centering
\includegraphics[width=0.49\textwidth]{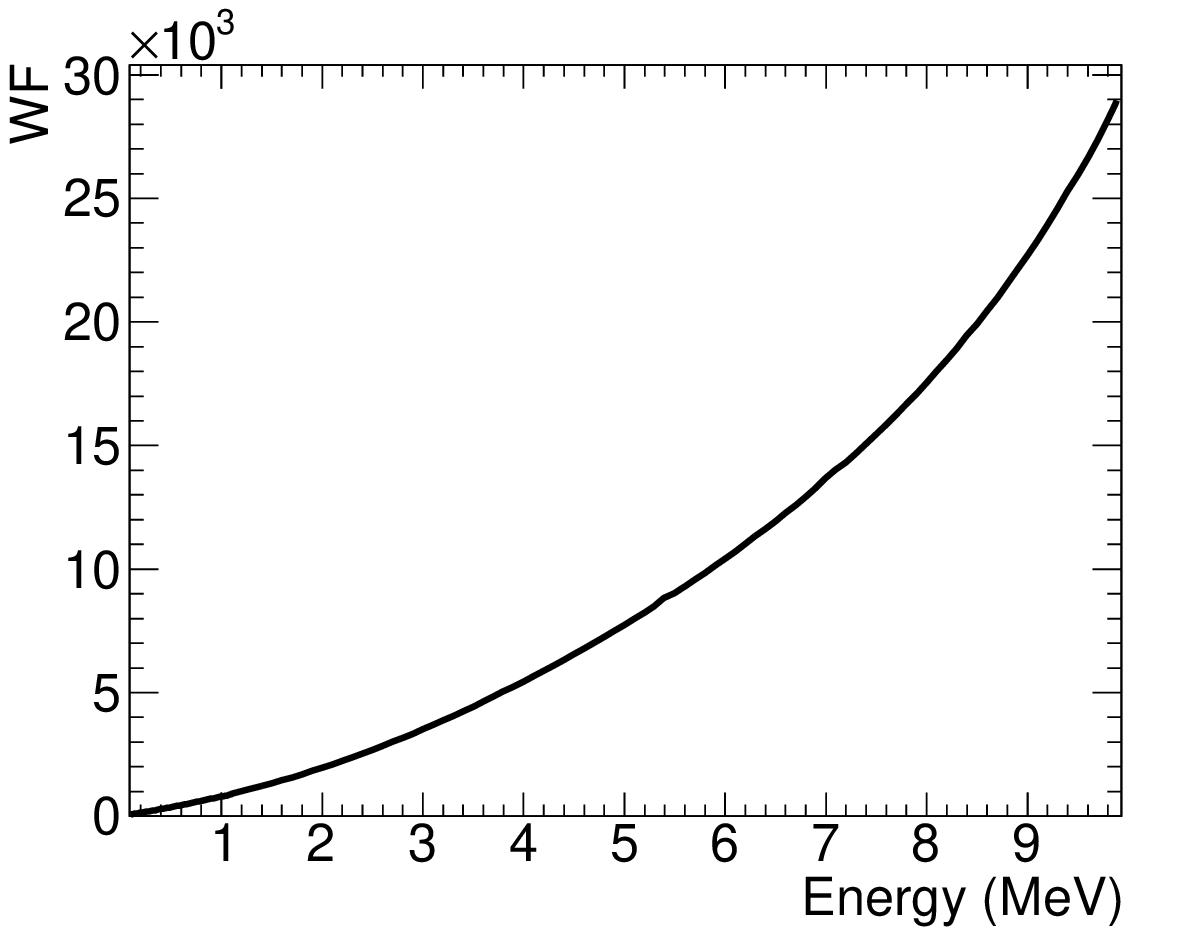}
\caption{Adopted WF for one sTED module. WF$\mathrm{(E)}$ = $\mathrm{6.5071}$ + $\mathrm{656.261\cdot E}$ + $\mathrm{113.929\cdot E^2}$ + $\mathrm{34.5488\cdot E^3}$-$\mathrm{6.39125\cdot E^4}$ + $\mathrm{0.411521\cdot E^5}$, the energy ($E$) is given in MeV and the coefficients correspond to $k=1$, see \ref{section:MeasuringTechnique}.}
\label{fig:WF} 
\end{figure}

In order to apply the PHWT, the WF has to be calculated for the actual detection setup. As described in reference \cite{Tain_04}, the best known method for determining the WF is to use accurate Monte Carlo simulations including the detailed geometrical description of the full experimental setup. For this reason, in Section \ref{section:DetectorResponse} the Monte Carlo simulations of an sTED module have been thoroughly validated.
The WF for a setup consisting of three sTED modules has been calculated. This particular setup was utilized to determine the capture yield of $^{197}$Au in EAR2, as explained in Section \ref{section:Validation}. The detector response was obtained for many $\gamma$-ray energies using the geometry depicted in Figure \ref{fig:Geant4Setup}. The simulation results were then fitted to derive a WF parameterized as a 5th-degree polynomial for each sTED module, following the procedure described in \cite{Tain_04}. The WF for one of the modules can be seen in Fig. \ref{fig:WF}. The WFs for the other two modules are very similar.


\begin{figure}[!ht]
\centering
\includegraphics[width=0.49\textwidth]{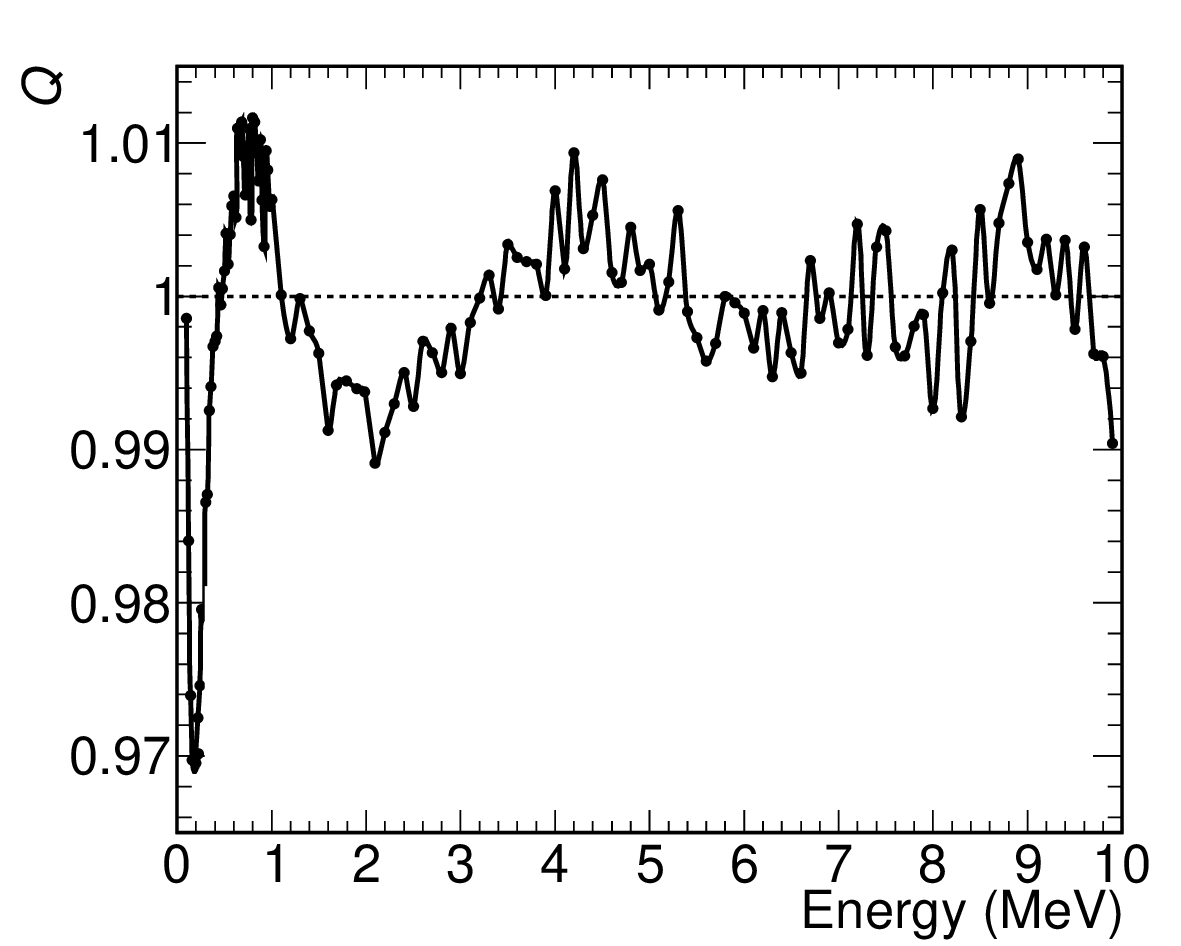}
\caption{$Q$ values, see Sec. \ref{section:MeasuringTechnique} for its definition, determined for 150 $\gamma$-ray energies.}
\label{fig:CheckWF} 
\end{figure}

A polynomial form is used to parameterize the WF as a function of the deposited energy. For each of 150 $\gamma$-ray energies, covering the range from 0.1 to 10 MeV, we simulated the spectrum of deposited energies. The detection efficiency ($\varepsilon_{\gamma \; (with \; WF) }$) for each $\gamma$-ray energy (E$_\gamma$) was then determined from these spectra weighted with the WF of Fig. \ref{fig:WF} and the obtained efficiency was divided by the corresponding $E_\gamma$ to determine the value $Q$. This quantity is defined in a mathematical equation as follows: Q=$\varepsilon_{\gamma \; (with \; WF)}$/$E_\gamma$. The appropriate WF should yield $Q=k$ ($k \equiv 1$ in our case). The $Q$ values obtained for this WF are shown in Fig. \ref{fig:CheckWF}, the small deviations from one demonstrate the high accuracy reached with the 5th degree polynomial WF. At energies below 0.5 MeV, the probability of absorption of all the energy of the $\gamma$-rays increases considerably, modifying the detector response, however, the determined WF is still able to obtain $Q$ values at this energy with a deviation of less than 3\%. 
These small deviations in the WFs have lead to uncertainties in the detection efficiency of capture $\gamma$-ray cascades of the order of only 0.3\% for the majority of the isotopes, similar values have been reported in previous works \cite{Tain_04,Alcayne_22,Lerendegui_18}.

 \section{Experimental validation at n\_TOF EAR2}\label{section:Validation}
 
\begin{figure}[!ht]
\centering
\includegraphics[width=0.49\textwidth]{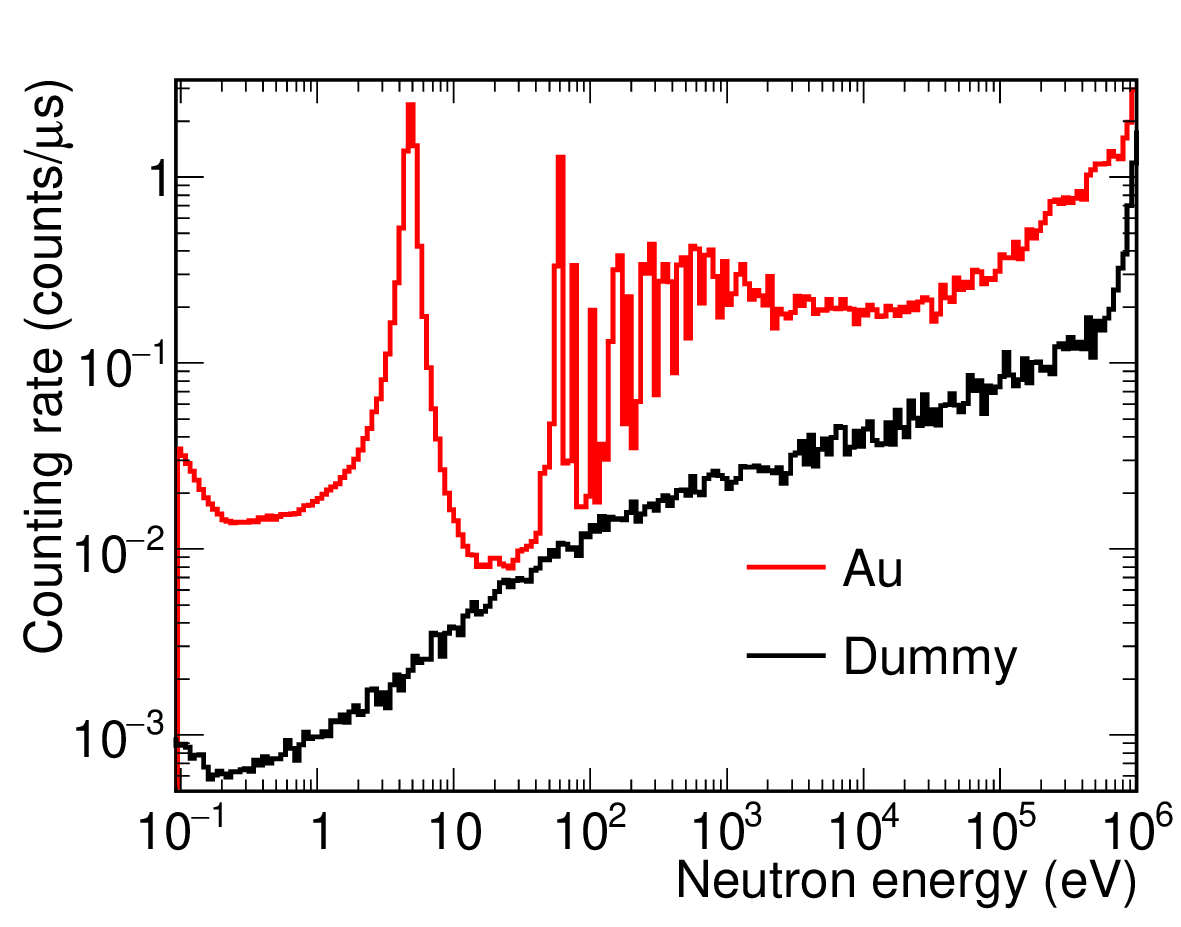}
\caption{Counting rates obtained as a function of the neutron energy in the experimental EAR2 for a sTED module with a threshold of 0.15 MeV. The detector is located at 5 cm from a $^{197}$Au sample of 2 cm in diameter and 100 $\mu$m thickness. The counting rates of the background obtained when measuring a dummy sample are also presented.}
\label{fig:CR_sTED} 
\end{figure}

Due to the challenging conditions for performing capture measurements at n\_TOF EAR2 described in Section \ref{section:Limitations}, the most reasonable method to validate the sTED is to perform a capture experiment and compare the results with the evaluated cross section data. A suitable isotope for this purpose is $^{197}$Au, which can be obtained in the form of high-purity metallic samples and has a standard capture cross-section at thermal energy and between 0.2 and 2.5 MeV \cite{Carlson_09}. In addition, $^{197}$Au has been measured many times at n\_TOF as a reference or in dedicated campaigns \cite{Massimi_10,Lederer_Au_11}. In 2022, a capture measurement was carried out with three sTED modules placed horizontally at 5 cm from the center of a 2 cm diameter and 0.1 mm thickness $^{197}$Au sample, see Fig. \ref{fig:Geant4Setup}. As presented in Fig. \ref{fig:CR_sTED} the counting rate obtained in this measurement with a sTED module is considerably lower than the one obtained with a BICRON detector at the same experimental area, see Fig. \ref{fig:CR_EAR1_EAR2}.

\begin{figure}[!ht]
\centering
\includegraphics[width=0.49\textwidth]{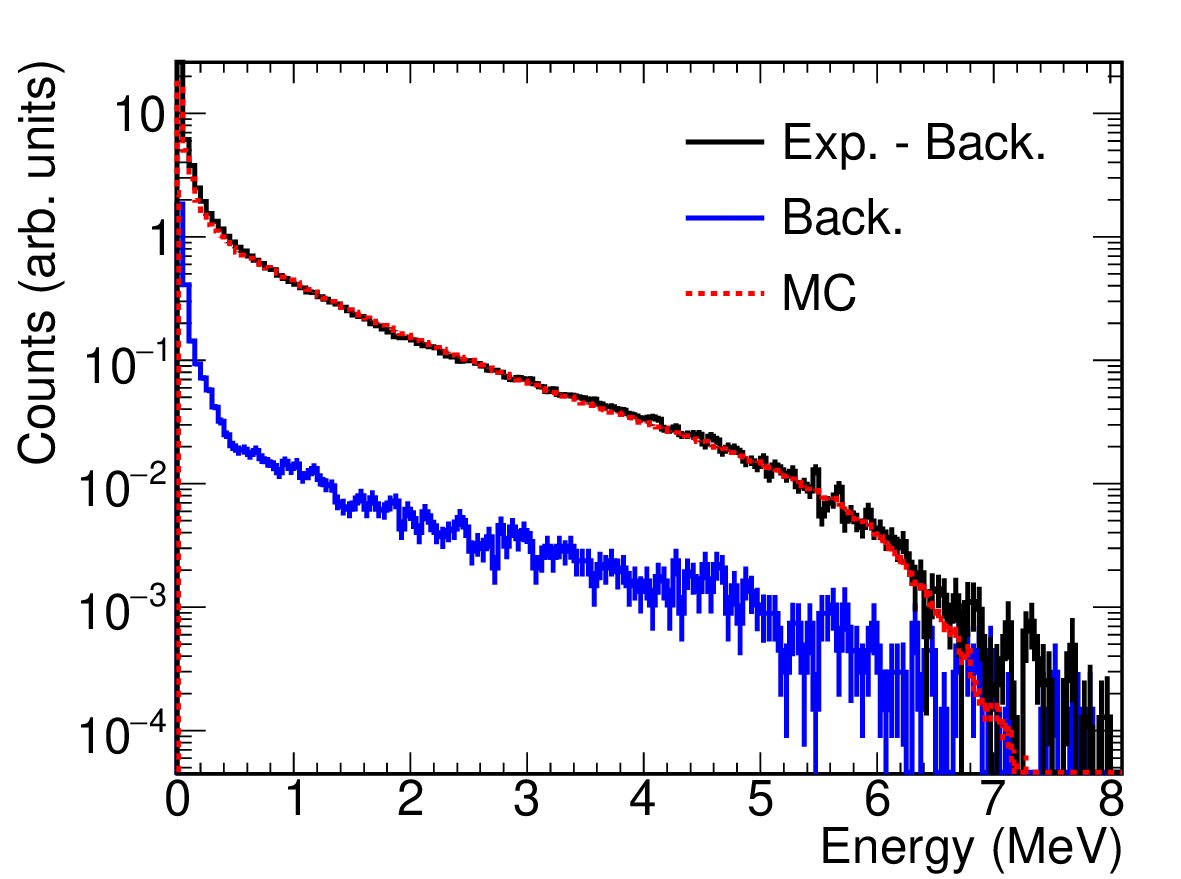}
\caption{Experimental deposited energy spectra in one sTED module (Exp.-Back.) with background (Back.) subtracted and simulated with Geant4 (MC) for $^{197}$Au (n,$\gamma$) cascades.}
\label{fig:AuEdep} 
\end{figure}

The deposited energy spectra measured for (n,$\gamma$) reactions in $^{197}$Au at EAR2 were compared with simulations performed with Geant4 in Fig. \ref{fig:AuEdep}. The $\gamma$-cascades used in the simulations were obtained by fitting the Total Absorption Calorimeter (TAC) data from a measurement performed at EAR1 \cite{Guerrero_09,Guerrero_12} and have been used in processing data of other experiments\cite{Alcayne_22,Mendoza_14,Mendoza_18}. The agreement between the shape of the experiment and simulations indicates that there are no significant gain shifts in the detector with respect to calibrations performed with sources.

\begin{figure}[!ht]
\centering
\includegraphics[width=0.49\textwidth]{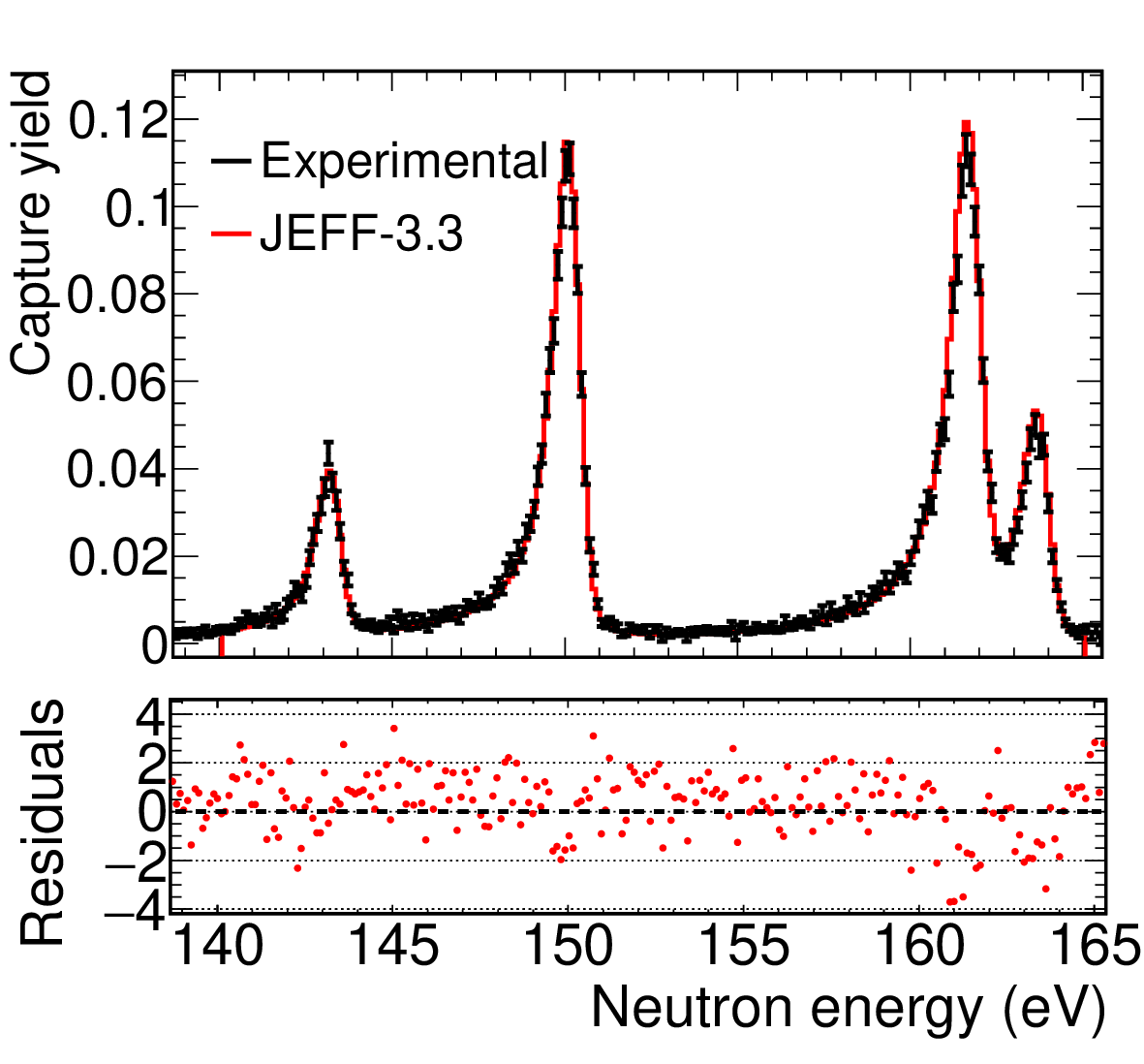}
\caption{sTED experimental capture yield obtained with a $^{197}$Au sample (Experimental) compared with the yield obtained from the JEFF-3.3 nuclear data library (JEFF-3.3) in the energy region between 140 and 165 eV. In the bottom panel of the figure, the residuals defined as the distances of the experimental data points to the theoretical JEFF-3.3 yield divided by the statistical uncertainties of the data points are plotted. The error bars consider only the uncertainties due to counting statistics.}
\label{fig:YieldResiduals} 
\end{figure}

\begin{figure}[!ht]
\centering
\includegraphics[width=0.49\textwidth]{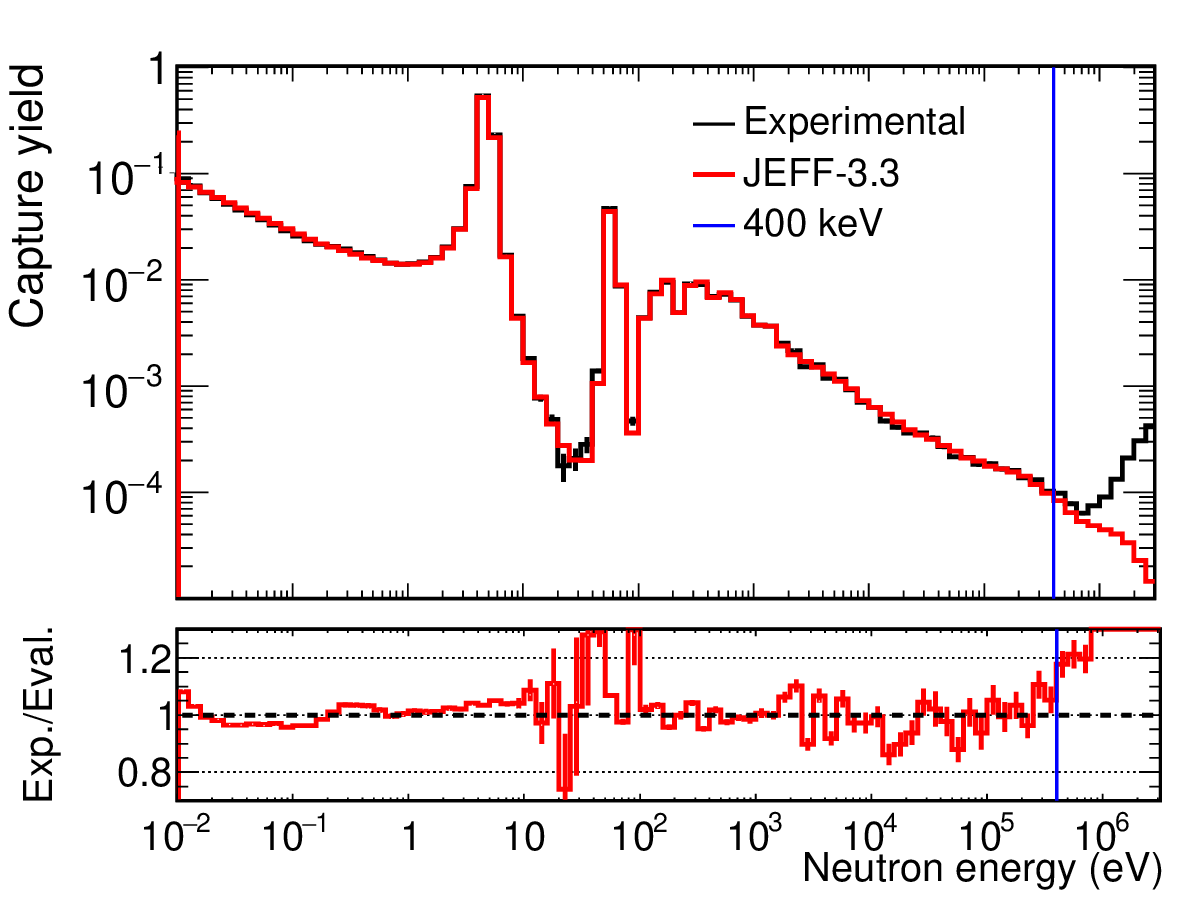} 
\includegraphics[width=0.49\textwidth]{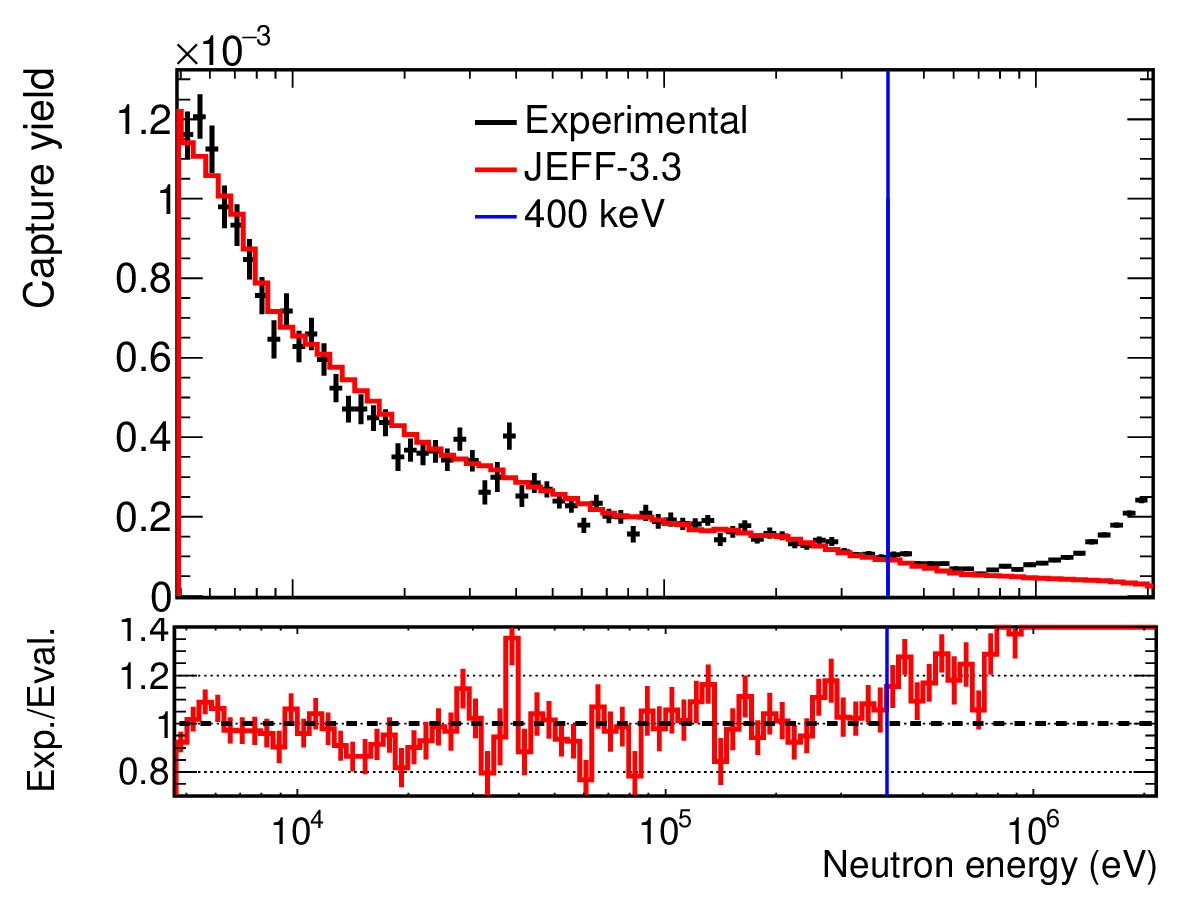}
\caption{sTED experimental capture yield obtained with a $^{197}$Au sample (Experimental) compared with the yield obtained from the JEFF-3.3 nuclear data library (JEFF-3.3). The top figure has ten bins per decade and the bottom one thirty bins per decade. The vertical blue line indicates the neutron energy of 400 keV. In the bottom panels, the ratios between the two yields are presented. The error bars consider only the uncertainties due to counting statistics.}
\label{fig:YieldRatio} 
\end{figure}

The capture yield has been determined by applying the WF calculated in Section \ref{section:MeasuringTechnique} to obtain the weighted counting rate as a function of the neutron energy and dividing it by the neutron fluence of the EAR2, after subtracting the different background components. The experimental yield was compared with the yield obtained with the JEFF-3.3 \cite{JEFF3.3_17} cross-section broadened with the Resolution Function (RF) of the EAR2 \cite{Vlachoudis_21,Pavon_22}. The comparison, normalized in the region between 0.01 and 1 eV, is presented in Figs. \ref{fig:YieldResiduals} and \ref{fig:YieldRatio}. As can be observed, the yields are very similar below 400 keV, which correspond to neutrons that are at least at $\sim$2 $\mu$s from the strong EAR2 particle flash. The small differences observed between our results and JEFF-3.3 can be attributed to uncertainties in the preliminary fluence shape and/or in the RF \cite{Pavon_22}. These quantities are preliminary at the moment because the spallation target has been changed recently at n\_TOF. In the valleys of the resonances in the energy range between 10 and 100 eV, the differences are larger than 25\% due to the uncertainties in the background subtraction. At energies higher than 400 keV, there are considerable differences, which are likely attributed to the opening of the (n,n$^{\shortmid}$) inelastic reaction channels at $\sim$100 keV, which have not been considered in the analysis.

The main conclusion of the analysis is that the sTED detector is capable of measuring a capture cross section up to at least 400 keV without suffering any degradation of its performance and thus are an excellent tool for measurements of the capture cross section using the high intensity beam of EAR2. Also, the detector is a very good option for any other possible facility facing high counting rates in capture cross section measurements.

The sTED detector has already been used at n\_TOF EAR2 to perform capture measurements of several isotopes ($^{79}$Se, $^{94}$Nb, $^{160}$Gd and $^{94,95,96}$Mo) in various geometric configurations, producing very promising data that are going to be published soon \cite{Lerendegui_23,Domingo_22,Balibrea_94Nb_23,Mastromarco_Gd_23,Mucciola_Mo_23}.

\section{Summary and conclusions}\label{section:Conclusions}

The performance of previously used C$_6$D$_6$ detectors at n\_TOF EAR2 harsh conditions for capture measurements has been discussed showing many limitations. An alternative segmented Total Energy Detector (sTED), also based on the C$_6$D$_6$ liquid scintillator, has been developed for improving the response of the previously-used detectors to the high counting rates. The sTED detector system will consist of an array of smaller active volume modules coupled to photo-multipliers optimized for high counting rates. The main features of the sTED to perform capture cross section measurements are:

\begin{itemize}
 \item The detector shows a linear energy response to $\gamma$-rays in the entire energy range considered.
 \item The experimental response of the sTED modules to $\gamma$-ray sources is well reproduced with Monte Carlo simulations. This is important to calibrate the detector and for the application of the PHWT.
 \item The applicability of the PHWT is validated by Monte Carlo simulations, by verifying that a WF is capable of producing a weighted efficiency proportional to the $\gamma$-ray energy.

\end{itemize}

 Last, but not least, an experimental campaign was carried out for validating the sTED performance under the demanding conditions of the n\_TOF EAR2. The experimental capture yield obtained for a $^{197}$Au sample was compared to predictions based on the JEFF-3.3 capture cross-section, showing an excellent agreement. The data show that the detector is capable of measuring a capture cross-section at EAR2 up to at least 400 keV, which is far above the limit reached with large volume C$_6$D$_6$. The sTED has been used already in several experimental campaigns producing very promising data.
\section*{CRediT authorship contribution statement}
This is a scientific article made by the n\_TOF collaboration, where its members have contributed to all aspects of its development.

\section*{Declaration of competing interest}

Declaration of competing interest The authors declare that they have no known competing financial interests or personal relationships that could have appeared to influence the work reported in this paper

\section*{Acknowledgments}

This work was supported in part by the I+D+i grant PGC2018-096717-B-C21 funded by MCIN/AEI/10.13039/501100011033, by project PID2021-123100NB-I00 funded by  MCIN/AEI/10.13039/501100011033/FEDER, UE, by project PCI2022-135037-2 funded by MCIN/AEI/10.13039/501100011033/ and  European Union NextGenerationEU/PRTR, by the European Commission H2020 Framework Programme project SANDA (Grant agreement ID: 847552) and by funding agencies of the n\_TOF participating institutions.\\

\bibliography{sTEDBiblio}

\begin{thebibliography}{46}
\expandafter\ifx\csname natexlab\endcsname\relax\def\natexlab#1{#1}\fi
\providecommand{\url}[1]{\texttt{#1}}
\providecommand{\href}[2]{#2}
\providecommand{\path}[1]{#1}
\providecommand{\DOIprefix}{doi:}
\providecommand{\ArXivprefix}{arXiv:}
\providecommand{\URLprefix}{URL: }
\providecommand{\Pubmedprefix}{pmid:}
\providecommand{\doi}[1]{\href{http://dx.doi.org/#1}{\path{#1}}}
\providecommand{\Pubmed}[1]{\href{pmid:#1}{\path{#1}}}
\providecommand{\bibinfo}[2]{#2}
\ifx\xfnm\relax \def\xfnm[#1]{\unskip,\space#1}\fi
\bibitem[{Esposito et~al.(2021)}]{Esposito_target_23}
\bibinfo{author}{R.~Esposito}, et~al. (\bibinfo{collaboration}{for the n\_TOF
  Collaboration}), \bibinfo{journal}{Phys. Rev. Accel. Beams}
  \bibinfo{volume}{24} (\bibinfo{year}{2021}) \bibinfo{pages}{093001}.
  \DOIprefix\doi{10.1103/PhysRevAccelBeams.24.093001}.
\bibitem[{Guerrero et~al.(2013)}]{Guerrero_13}
\bibinfo{author}{C.~Guerrero}, et~al. (\bibinfo{collaboration}{The n\_TOF
  Collaboration}), \bibinfo{journal}{Eur. Phys. J. A} \bibinfo{volume}{49}
  (\bibinfo{year}{2013}) \bibinfo{pages}{27}.
  \DOIprefix\doi{10.1140/epja/i2013-13027-6}.
\bibitem[{Weiss et~al.(2015)}]{Weiss_15}
\bibinfo{author}{C.~Weiss}, et~al. (\bibinfo{collaboration}{The n\_TOF
  Collaboration}), \bibinfo{journal}{Nucl. Instrum. Methods A}
  \bibinfo{volume}{799} (\bibinfo{year}{2015}) \bibinfo{pages}{90--98}.
  \DOIprefix\doi{10.1016/j.nima.2015.07.027}.
\bibitem[{Ferrari et~al.(2022)}]{Ferrarri_23}
\bibinfo{author}{M.~Ferrari}, et~al., \bibinfo{journal}{Phys. Rev. Ac. and Be.}
  \bibinfo{volume}{25} (\bibinfo{year}{2022}).
  \DOIprefix\doi{10.1103/PhysRevAccelBeams.25.103001}.
\bibitem[{Gervino et~al.(2022)}]{Gervino_22}
\bibinfo{author}{Gervino}, et~al., \bibinfo{journal}{Universe}
  \bibinfo{volume}{8} (\bibinfo{year}{2022}). \URLprefix
  \url{https://www.mdpi.com/2218-1997/8/5/255}.
  \DOIprefix\doi{10.3390/universe8050255}.
\bibitem[{Barbagallo et~al.(2016)}]{Barbagallo_16}
\bibinfo{author}{M.~Barbagallo}, et~al. (\bibinfo{collaboration}{The n\_TOF
  Collaboration}), \bibinfo{journal}{Phys. Rev. Lett.} \bibinfo{volume}{117}
  (\bibinfo{year}{2016}) \bibinfo{pages}{152701}.
  \DOIprefix\doi{10.1103/PhysRevLett.117.152701}.
\bibitem[{Sabat{\'e}-Gilarte et~al.(2017)}]{Sabate_17}
\bibinfo{author}{Sabat{\'e}-Gilarte}, et~al., \bibinfo{journal}{EPJ Web Conf.}
  \bibinfo{volume}{146} (\bibinfo{year}{2017}) \bibinfo{pages}{08004}.
  \DOIprefix\doi{10.1051/epjconf/201714608004}.
\bibitem[{Damone et~al.(2018)}]{Damone_18}
\bibinfo{author}{L.~Damone}, et~al. (\bibinfo{collaboration}{The n\_TOF
  Collaboration}), \bibinfo{journal}{Phys. Rev. Lett.} \bibinfo{volume}{121}
  (\bibinfo{year}{2018}) \bibinfo{pages}{042701}.
  \DOIprefix\doi{10.1103/PhysRevLett.121.042701}.
\bibitem[{Stamatopoulos et~al.(2020)}]{Stamatopoulos_20}
\bibinfo{author}{A.~Stamatopoulos}, et~al. (\bibinfo{collaboration}{n\_TOF
  Collaboration}), \bibinfo{journal}{Phys. Rev. C} \bibinfo{volume}{102}
  (\bibinfo{year}{2020}) \bibinfo{pages}{014616}.
  \DOIprefix\doi{10.1103/PhysRevC.102.014616}.
\bibitem[{Alcayne et~al.(2020)}]{Alcayne_19}
\bibinfo{author}{V.~Alcayne}, et~al., \bibinfo{journal}{EPJ Web Conf.}
  \bibinfo{volume}{239} (\bibinfo{year}{2020}) \bibinfo{pages}{01034}.
  \DOIprefix\doi{10.1051/epjconf/202023901034}.
\bibitem[{Guerrero et~al.(2014)}]{Guerrero_14}
\bibinfo{author}{C.~Guerrero}, et~al. (\bibinfo{collaboration}{The n\_TOF
  Collaboration}), \bibinfo{journal}{Nuclear Data Sheets} \bibinfo{volume}{119}
  (\bibinfo{year}{2014}) \bibinfo{pages}{5--9}.
  \DOIprefix\doi{https://doi.org/10.1016/j.nds.2014.08.004}.
\bibitem[{Gunsing et~al.(2017)}]{Gunsinsg_17}
\bibinfo{author}{F.~Gunsing}, et~al. (\bibinfo{collaboration}{The n\_TOF
  Collaboration}), \bibinfo{journal}{EPJ Web Conf.} \bibinfo{volume}{146}
  (\bibinfo{year}{2017}) \bibinfo{pages}{11002}.
  \DOIprefix\doi{10.1051/epjconf/201714611002}.
\bibitem[{Macklin and Gibbons(1967)}]{Macklin_67}
\bibinfo{author}{R.~L. Macklin}, \bibinfo{author}{J.~H. Gibbons},
  \bibinfo{journal}{Phys. Rev. C} \bibinfo{volume}{159} (\bibinfo{year}{1967})
  \bibinfo{pages}{1007}. \DOIprefix\doi{10.1103/PhysRev.159.1007}.
\bibitem[{Abbondanno et~al.(2004)}]{Tain_04}
\bibinfo{author}{U.~Abbondanno}, et~al. (\bibinfo{collaboration}{The n\_TOF
  Collaboration}), \bibinfo{journal}{Nucl. Instrum. Methods A}
  \bibinfo{volume}{521} (\bibinfo{year}{2004}) \bibinfo{pages}{454--467}.
  \DOIprefix\doi{10.1016/j.nima.2003.09.066}.
\bibitem[{Borella et~al.(2007)}]{Borella_07}
\bibinfo{author}{A.~Borella}, et~al., \bibinfo{journal}{Nucl. Instrum. Methods
  A} \bibinfo{volume}{577} (\bibinfo{year}{2007}) \bibinfo{pages}{626--640}.
  \DOIprefix\doi{10.1016/j.nima.2007.03.034}.
\bibitem[{Moxon and Rae(1963)}]{Moxon_63}
\bibinfo{author}{M.~Moxon}, \bibinfo{author}{E.~Rae}, \bibinfo{journal}{Nucl.
  Instrum. Methods} \bibinfo{volume}{24} (\bibinfo{year}{1963})
  \bibinfo{pages}{445--455}. \DOIprefix\doi{10.1016/0029-554X(63)90364-1}.
\bibitem[{Plag et~al.(2003)}]{Plag_03}
\bibinfo{author}{R.~Plag}, et~al., \bibinfo{journal}{Nucl. Instrum. Methods A}
  \bibinfo{volume}{496} (\bibinfo{year}{2003}) \bibinfo{pages}{425--436}.
  \DOIprefix\doi{10.1016/s0168-9002(02)01749-7}.
\bibitem[{Mastinu et~al.(2013)}]{Mastinu_13}
\bibinfo{author}{P.~Mastinu}, et~al., \bibinfo{title}{{New C6D6 detectors:
  reduced neutron sensitivity and improved safety}}, \bibinfo{type}{Technical
  Report}, \bibinfo{year}{2013}. \URLprefix
  \url{https://cdsweb.cern.ch/record/1558147/}.
\bibitem[{Oprea et~al.(2020)}]{Oprea_20}
\bibinfo{author}{A.~Oprea}, et~al., \bibinfo{journal}{EPJ Web Conf.}
  \bibinfo{volume}{239} (\bibinfo{year}{2020}) \bibinfo{pages}{01009}.
  \DOIprefix\doi{10.1051/epjconf/202023901009}.
\bibitem[{Alcayne(2022)}]{Alcayne_22}
\bibinfo{author}{V.~Alcayne}, \bibinfo{title}{{Measurement of the 244Cm, 246Cm
  and 248Cm neutron-induced capture cross sections at the CERN n TOF
  facility}}, Ph.D. thesis, \bibinfo{year}{2022}. \URLprefix
  \url{https://cds.cern.ch/record/2811791}.
\bibitem[{Alcayne et~al.(2023)}]{Alcayne_23}
\bibinfo{author}{V.~Alcayne}, et~al., \bibinfo{journal}{EPJ Web of Conf.}
  \bibinfo{volume}{284} (\bibinfo{year}{2023}) \bibinfo{pages}{01043}.
  \DOIprefix\doi{10.1051/epjconf/202328401043}.
\bibitem[{Guerrero et~al.(2008)}]{Guerrero_08}
\bibinfo{author}{C.~Guerrero}, et~al., \bibinfo{journal}{Nucl. Instrum. Methods
  A} \bibinfo{volume}{597} (\bibinfo{year}{2008}) \bibinfo{pages}{212--218}.
  \DOIprefix\doi{10.1016/j.nima.2008.09.017}.
\bibitem[{Mendoza et~al.(2023)}]{Mendoza_23}
\bibinfo{author}{E.~Mendoza}, et~al., \bibinfo{journal}{Nucl. Instrum. Methods
  A}  (\bibinfo{year}{2023}) \bibinfo{pages}{167894}.
  \DOIprefix\doi{10.1016/j.nima.2022.167894}.
\bibitem[{Sci(2022)}]{Scionix}
\bibinfo{title}{{Scionix}}, \bibinfo{year}{2022}. \URLprefix
  \url{https://scionix.nl}.
\bibitem[{Ham(2022)}]{Hamamatsu}
\bibinfo{title}{{Hamamatsu}}, \bibinfo{year}{2022}. \URLprefix
  \url{https://www.hamamatsu.com/}.
\bibitem[{Knoll(1979)}]{Knoll_79}
\bibinfo{author}{G.~Knoll}, \bibinfo{title}{{Radiation Detection and
  Measurement}}, number \bibinfo{number}{v. 415} in \bibinfo{series}{{Radiation
  Detection and Measurement}}, \bibinfo{publisher}{Wiley},
  \bibinfo{year}{1979}.
\bibitem[{{\v Z}ugec et~al.(2016)}]{Zugec_16}
\bibinfo{author}{{\v Z}ugec}, et~al. (\bibinfo{collaboration}{The n\_TOF
  Collaboration}), \bibinfo{journal}{Nucl. Instrum. Methods A}
  \bibinfo{volume}{812} (\bibinfo{year}{2016}) \bibinfo{pages}{134--144}.
  \DOIprefix\doi{10.1016/j.nima.2015.12.054}.
\bibitem[{Gramage(2016)}]{Gramage_16}
\bibinfo{author}{P.~Gramage}, \bibinfo{title}{{Characterization of large-scale
  LaBr 3 detectors volume for gamma spectroscopy}}, Master's thesis,
  \bibinfo{year}{2016}.
\bibitem[{Agostinelli et~al.(2003)}]{Agostinelli_03}
\bibinfo{author}{S.~Agostinelli}, et~al. (\bibinfo{collaboration}{the GEANT4
  Collaboration}), \bibinfo{journal}{Nucl. Instrum. Methods A}
  \bibinfo{volume}{506} (\bibinfo{year}{2003}) \bibinfo{pages}{250--303}.
  \DOIprefix\doi{10.1016/S0168-9002(03)01368-8}.
\bibitem[{Mendoza et~al.(2020)}]{Mendoza_20}
\bibinfo{author}{E.~Mendoza}, et~al., \bibinfo{journal}{EPJ Web Conf.}
  \bibinfo{volume}{239} (\bibinfo{year}{2020}) \bibinfo{pages}{17006}.
  \DOIprefix\doi{10.1051/epjconf/202023917006}.
\bibitem[{Lerendegui-Marco et~al.(2018)}]{Lerendegui_18}
\bibinfo{author}{J.~Lerendegui-Marco}, et~al. (\bibinfo{collaboration}{The
  n\_TOF Collaboration}), \bibinfo{journal}{Phys. Rev. C} \bibinfo{volume}{97}
  (\bibinfo{year}{2018}) \bibinfo{pages}{024605}.
  \DOIprefix\doi{10.1103/PhysRevC.97.024605}.
\bibitem[{Carlson et~al.(2009)}]{Carlson_09}
\bibinfo{author}{A.~Carlson}, et~al., \bibinfo{journal}{Nucl. Data Sheets}
  \bibinfo{volume}{110} (\bibinfo{year}{2009}) \bibinfo{pages}{3215--3324}.
  \DOIprefix\doi{10.1016/j.nds.2009.11.001}, \bibinfo{note}{special Issue on
  Nuclear Reaction Data}.
\bibitem[{Massimi et~al.(2010)}]{Massimi_10}
\bibinfo{author}{C.~Massimi}, et~al. (\bibinfo{collaboration}{n\_TOF
  Collaboration}), \bibinfo{journal}{Phys. Rev. C} \bibinfo{volume}{81}
  (\bibinfo{year}{2010}) \bibinfo{pages}{044616}.
  \DOIprefix\doi{10.1103/PhysRevC.81.044616}.
\bibitem[{Lederer et~al.(2011)}]{Lederer_Au_11}
\bibinfo{author}{C.~Lederer}, et~al. (\bibinfo{collaboration}{n\_TOF
  Collaboration}), \bibinfo{journal}{Phys. Rev. C} \bibinfo{volume}{83}
  (\bibinfo{year}{2011}) \bibinfo{pages}{034608}.
  \DOIprefix\doi{10.1103/PhysRevC.83.034608}.
\bibitem[{Guerrero et~al.(2009)}]{Guerrero_09}
\bibinfo{author}{C.~Guerrero}, et~al., \bibinfo{journal}{Nucl. Instrum. Methods
  A} \bibinfo{volume}{608} (\bibinfo{year}{2009}) \bibinfo{pages}{424--433}.
  \DOIprefix\doi{10.1016/j.nima.2009.07.025}.
\bibitem[{Guerrero et~al.(2012)}]{Guerrero_12}
\bibinfo{author}{C.~Guerrero}, et~al., \bibinfo{journal}{Nucl. Instrum. Methods
  A} \bibinfo{volume}{671} (\bibinfo{year}{2012}) \bibinfo{pages}{108--117}.
  \DOIprefix\doi{10.1016/j.nima.2011.12.046}.
\bibitem[{Mendoza et~al.(2014)}]{Mendoza_14}
\bibinfo{author}{E.~Mendoza}, et~al. (\bibinfo{collaboration}{n\_TOF
  Collaboration}), \bibinfo{journal}{Phys. Rev. C} \bibinfo{volume}{90}
  (\bibinfo{year}{2014}) \bibinfo{pages}{034608}.
  \DOIprefix\doi{10.1103/PhysRevC.90.034608}.
\bibitem[{Mendoza et~al.(2018)}]{Mendoza_18}
\bibinfo{author}{E.~Mendoza}, et~al. (\bibinfo{collaboration}{n\_TOF
  Collaboration}), \bibinfo{journal}{Phys. Rev. C} \bibinfo{volume}{97}
  (\bibinfo{year}{2018}) \bibinfo{pages}{054616}.
  \DOIprefix\doi{10.1103/PhysRevC.97.054616}.
\bibitem[{Plompen et~al.(2020)}]{JEFF3.3_17}
\bibinfo{author}{A.~J.~M. Plompen}, et~al., \bibinfo{journal}{Eur. Phys. J. A}
  \bibinfo{volume}{56} (\bibinfo{year}{2020}) \bibinfo{pages}{181}.
  \DOIprefix\doi{10.1140/epja/s10050-020-00141-9}.
\bibitem[{Vlachoudis et~al.(2021)}]{Vlachoudis_21}
\bibinfo{author}{V.~Vlachoudis}, et~al., \bibinfo{title}{{On the resolution
  function of the n\_TOF facility: a comprehensive study and user guide}},
  \bibinfo{type}{Technical Report}, \bibinfo{year}{2021}. \URLprefix
  \url{https://cds.cern.ch/record/2764434}.
\bibitem[{{Pav\'on-Rodr\'{\i}guez} et~al.(2023)}]{Pavon_22}
\bibinfo{author}{J.~A. {Pav\'on-Rodr\'{\i}guez}}, et~al., \bibinfo{journal}{EPJ
  Web of Conf.} \bibinfo{volume}{284} (\bibinfo{year}{2023})
  \bibinfo{pages}{06006}. \DOIprefix\doi{10.1051/epjconf/202328406006}.
\bibitem[{Lerendegui-Marco et~al.(2023)}]{Lerendegui_23}
\bibinfo{author}{J.~Lerendegui-Marco}, et~al., \bibinfo{journal}{EPJ Web of
  Conf.} \bibinfo{volume}{284} (\bibinfo{year}{2023}) \bibinfo{pages}{01028}.
  \DOIprefix\doi{10.1051/epjconf/202328401028}.
\bibitem[{Domingo-Pardo et~al.(2023)}]{Domingo_22}
\bibinfo{author}{C.~Domingo-Pardo}, et~al., \bibinfo{journal}{Eur. Phys. J. A}
  \bibinfo{volume}{59} (\bibinfo{year}{2023}) \bibinfo{pages}{8}.
  \DOIprefix\doi{10.1140/epja/s10050-022-00876-7}.
\bibitem[{Balibrea-Correa et~al.(2023)}]{Balibrea_94Nb_23}
\bibinfo{author}{J.~Balibrea-Correa}, et~al., \bibinfo{journal}{EPJ Web Conf.}
  \bibinfo{volume}{279} (\bibinfo{year}{2023}) \bibinfo{pages}{06004}.
  \DOIprefix\doi{10.1051/epjconf/202327906004}.
\bibitem[{Mastromarco et~al.(2023)}]{Mastromarco_Gd_23}
\bibinfo{author}{M.~Mastromarco}, et~al., \bibinfo{journal}{EPJ Web of Conf.}
  \bibinfo{volume}{284} (\bibinfo{year}{2023}) \bibinfo{pages}{09002}.
  \DOIprefix\doi{10.1051/epjconf/202328409002}.
\bibitem[{Mucciola et~al.(2023)}]{Mucciola_Mo_23}
\bibinfo{author}{R.~Mucciola}, et~al., \bibinfo{journal}{EPJ Web of Conf.}
  \bibinfo{volume}{284} (\bibinfo{year}{2023}) \bibinfo{pages}{01031}.
  \DOIprefix\doi{10.1051/epjconf/202328401031}.

\end{thebibliography}

\end{document}